\documentstyle[aps,eqsecnum,epsf]{revtex}

\ifx\epsffile\undefined
\def\insertfig#1{}
\else
\def\insertfig#1{{\baselineskip=4pt
\centerline{\epsfxsize=\hsize\epsffile{#1}}}}\fi

\begin{document}

\def\slash#1{\rlap{$#1$}/} 
\def\dsl{\,\raise.15ex\hbox{/}\mkern-13.5mu D} 
\def\delsl{\raise.15ex\hbox{/}\kern-.57em\partial}
\def\grad#1{\,\nabla\!^{{#1}}\,}
\def\det{{\rm det}}
\def\Tr{{\rm Tr}}
\def\N{N_c}
\def\L{{\cal L}}
\def\P{{\cal P}}
\def\S{{\cal S}}
\def\B{{\cal B}}
\def\D{{\cal D}}
\def\A{{\cal A}}
\def\M{{\cal M}}
\def\T{{\cal T}}
\def\H{{\cal H}}
\def\C{{\cal C}}
\def\V{{\cal V}}
\def\O{{\cal O}}
\def\ltap{\ \raise.3ex\hbox{$<$\kern-.75em\lower1ex\hbox{$\sim$}}\ }
\def\gtap{\ \raise.3ex\hbox{$>$\kern-.75em\lower1ex\hbox{$\sim$}}\ }
\def\ssqr#1#2{{\vbox{\hrule height #2pt
      \hbox{\vrule width #2pt height#1pt \kern#1pt\vrule width #2pt}
      \hrule height #2pt}\kern- #2pt}}
\def\sqr{\mathchoice\ssqr8{.4}\ssqr8{.4}\ssqr{5}{.3}\ssqr{4}{.3}}

\def\bsqr{\ssqr{10}{.1}}
\def\nbox{\hbox{$\bsqr\bsqr\bsqr\bsqr\raise2.7pt\hbox{$\,\cdot\cdot
\cdot\cdot\cdot\,$}\bsqr\bsqr\bsqr$}}
\def\onebox{{\vbox{\hbox{$\sqr\thinspace$}}}}
\def\twobox{{\vbox{\hbox{$\sqr\sqr\thinspace$}}}}
\def\threebox{{\vbox{\hbox{$\sqr\sqr\sqr\thinspace$}}}}

\twocolumn[       
{\tighten
\preprint{\vbox{
\hbox{UCSD/PTH 95--17}
\hbox{hep-ph/95xxxxx}
}}
\title{Chiral Lagrangian for Baryons in the 1/$\bf \N$ Expansion}
\author{Elizabeth Jenkins\footnotemark\footnotemark}
\address{Department of Physics, University of California at
San Diego, La Jolla, CA 92093}
\bigskip
\date{September 1995}
\maketitle
\widetext
\vskip-1.5in
\rightline{\vbox{
\hbox{UCSD/PTH 95--17}
\hbox{hep-ph/95xxxxx}
}}
\vskip1.5in
\begin{abstract}
A $1/\N$ expansion of the chiral Lagrangian for baryons
is formulated and used to study
the low-energy dynamics of baryons interacting with
the pion nonet $\pi$, $K$, $\eta$ and $\eta^\prime$
in a combined expansion in chiral symmetry breaking
and $1/\N$.  Strong $CP$-violation is included.
The chiral Lagrangian correctly implements nonet symmetry
and contracted spin-flavor symmetry for baryons
in the large $\N$ limit.
The implications of nonet
symmetry for low-energy baryon-pion interactions are
described in detail.  The procedure for calculating
non-analytic pion-loop corrections to baryon amplitudes
in the $1/\N$ expansion for finite $\N$ is explained.
Flavor-$\bf 27$ baryon mass splittings are calculated
at leading order in chiral perturbation theory as an
example.
\end{abstract}

}

] 

\narrowtext

\footnotetext{${}^\dagger$Alfred P. Sloan Research Fellow.}
\footnotetext{${}^*$National Science Foundation Young Investigator.}
\section{Introduction}

Although it is by now well-established that the theory
of the strong interactions is quantum chromodynamics,
first principles calculations of the
spectrum and properties of hadrons are not
possible
because the theory is strongly coupled at low energies.
A number of different methods have been used to extract
low-energy consequences of QCD.  One of the oldest
methods is chiral perturbation theory\cite{chpt}
which exploits the symmetry of the QCD Lagrangian
under $SU(3)_L \times SU(3)_R \times U(1)_V$ transformations
on the three flavors of light quarks $u$, $d$ and $s$
in the limit that the quark masses $m_u$, $m_d$ and $m_s$
vanish.  Chiral symmetry is spontaneously broken
to the vector subgroup $SU(3) \times U(1)_V$ by the QCD vacuum,
resulting in an octet of pseudoscalar Goldstone bosons,
the pions.
A perturbative expansion in the pion momenta and the
explicit chiral symmetry breaking parameters
$m_i$ over the scale parameter of
chiral symmetry breaking $\Lambda_\chi$ leads
to flavor symmetry relations amongst hadronic
amplitudes which are valid to a given order in chiral
symmetry breaking.

A second method which has been important in the understanding
of low-energy QCD hadron dynamics
is the $1/\N$ expansion \cite{thooft}.  This method promotes
QCD to a $SU(\N)$ non-abelian gauge theory, where $\N$ is
the number of colors.  The $1/\N$ expansion has been used
primarily to derive $1/\N$ power counting rules for hadronic
amplitudes \cite{thooft,witbaryons,coleman}.
For finite and large $\N$, planar
diagrams dominate the dynamics.  Each quark loop is suppressed
by one factor of $1/\N$ and non-planar gluon exchange is
suppressed by two factors of $1/\N$.  The suppression of quark
loops in the $1/\N$ expansion is
particularly important for processes involving hadrons, since it
implies that diagrams of
leading order in the $1/\N$ expansion contain no quark--anti-quark
pair creation and annihilation.  Thus,
planar QCD has a flavor symmetry \cite{planarqcd}
\begin{equation}\label{planarsym}
U(1)_{q_i} \times U(1)_{\overline q_i}
\end{equation}
which allows independent rotations on each quark flavor
and antiquark flavor and implies the separate conservation of the number
of each quark flavor and of each anti-quark flavor (light or heavy).
The planar QCD flavor symmetry (\ref{planarsym})
is broken at first subleading order due
to a single quark loop of order $1/\N$.
It is important to emphasize that (\ref{planarsym}) is a symmetry of
the planar approximation of QCD dynamics only and {\it not} of the
QCD Lagrangian itself.
Consequences of planar flavor symmetry include Zweig's rule and the
formation of ideally mixed meson nonets (in the $SU(3)$ flavor limit)
at leading order in $1/\N$ \cite{veneziano}.
Planar flavor symmetry is often called
``nonet symmetry" in the literature for this reason.

The combined use of chiral perturbation theory and the $1/\N$ expansion
can constrain the low-energy interactions of hadrons with the pion
nonet $\pi$, $K$, $\eta$ and $\eta^\prime$ more effectively
than either
method alone.  An effective Lagrangian describing the spectrum and
self-interactions of the pion nonet
was constructed some time ago \cite{dVV,witten,rst}.
The derivation of this $1/\N$ chiral Lagrangian
led to a number of
important theoretical results concerning the QCD vacuum angle
$\theta$, and to a consistent picture for phenomenology \cite{dVV,vnpv}
associated with the resolution of the
$U(1)_A$ problem \cite{uonea,wvv}.
Two results deserve special mention here.
First, the phenomenological analysis
proved that the $\theta$ parameter is close to zero in QCD
\cite{dVV,witten}.
Second, the analysis showed that an $\eta^\prime$ which
is primarily an $SU(3)$ flavor singlet (in violation of ideal mixing)
and which has a mass much larger than the pion octet is accommodated
for reasonable values of parameters\cite{dVV}.
Understanding these features of the $\eta^\prime$ is nontrivial
because the phenomenology involves
an interplay between effects suppressed by $m_i/\Lambda_\chi$ and $1/\N$.

In this work, a $1/\N$ chiral Lagrangian for the lowest-lying baryons
is constructed.  The Lagrangian
describes the interactions of the spin-$1/2$
baryon octet and the spin-$3/2$ baryon decuplet with the pion nonet.
The formulation of the $1/\N$ baryon chiral Lagrangian relies upon
recent developments in the study of the spin-flavor structure
of baryons in the $1/\N$ expansion
\cite{dm,j,djm,jm,djmtwo,jl,ddjm,cgo,cgkm,lm,luty,lmw,bron,manohar,matt,%
matt2,tsk}.
In the large-$\N$ limit, the baryon sector of QCD
possesses an exact contracted spin-flavor symmetry
algebra \cite{dm,gervaissakita}.  For finite $\N$,
corrections to the large-$\N$ limit are parametrized by $1/\N$-suppressed
operators \cite{dm,j}.
Consistency conditions determine which operators are allowed at
any given order in the $1/\N$ expansion.
The (spin $\otimes$ flavor) structure of the $1/\N$ expansion for baryons is
manifest in the baryon chiral Lagrangian presented here.
In addition, planar QCD flavor symmetry is implemented at leading
order in $1/\N$, and violated at first subleading order.  Planar QCD flavor
symmetry leads to a number of new results, such as the formation
of flavor nonets amongst baryon flavor octet and singlet amplitudes at
leading order in $1/\N$.
The consequences of planar QCD flavor symmetry are examined in detail, and
are entirely new to this work.
Planar QCD flavor symmetry follows from the
$1/\N$ expansion alone, so the results of this symmetry for
baryons do not depend on the chiral Lagrangian framework and
are valid in general.
Strong $CP$-violation enters the baryon chiral Lagrangian in the same
manner as earlier treatments \cite{divech,pich,baluni,crewther}.
Finally, the issue of non-analytic meson-loop
corrections to baryons amplitudes is examined.  A number of subtleties
arise in the calculation of loop corrections at finite $\N$.  The procedure
for computing loop calculations using operators at finite $\N$ is explained.
The group theoretic and $1/\N$ structure of these corrections is explicit
in this method.

The paper is organized as follows.  A presentation of
the pion nonet chiral Lagrangian
is given in Sect.~II to set notation.  Readers familiar with the
$1/N_c$ chiral Lagrangian for the pseudo-Goldstone bosons
may skip directly to Sect.~III and refer
to Sect.~II only for definitions of the meson nonet field $\bbox{\Phi}$
and strong-$CP$ parameters.
Sect.~III presents the $1/\N$ baryon chiral Lagrangian.  The baryon chiral
Lagrangian is formulated for arbitrary finite $\N$
in terms of the $1/\N$ operator expansion for baryons.  Planar QCD flavor
symmetry is imposed on the Lagrangian at leading order in $1/\N$.  The
$1/\N$ baryon chiral Lagrangian for $\N=3$ is compared to the
chiral Lagrangian for the spin-$1/2$ octet and spin-$3/2$ decuplet baryons
with no $1/\N$ expansion.  Sect.~IV addresses the computation of
non-analytic corrections using the $1/\N$ baryon chiral Lagrangian.  The
flavor-${\bf 27}$ non-analytic contribution to baryon masses
is computed to illustrate the method.  An understanding of the
accuracy of the Gell-Mann--Okubo formula for baryon octet masses is gained
from this computation.  Sect.~V considers the implications of $U(2)$ planar QCD
flavor symmetry for $SU(3)$ breaking of the baryon $1/\N$ expansion.
Conclusions are presented in Sect.~VI.

\section{Pion Chiral Lagrangian}
The $1/\N$ chiral Lagrangian describing the interactions of baryons and
low-momentum pions has the form
\begin{equation}
{\L} = {\L}_{\rm pion} + {\L}_{\rm baryon},
\end{equation}
where the pion Lagrangian describes the self-interactions of the
pseudo-Goldstone boson nonet.
In order to calculate in chiral perturbation theory to non-trivial orders
in the $1/\N$ expansion for baryons, it is necessary to understand the
$1/\N$ chiral Lagrangian of the pion sector
as well.  This section contains a presentation of the pion nonet
$1/\N$ chiral Lagrangian, as originally derived by
di Vecchia and Veneziano \cite{dVV} and
Witten \cite{witten}.  The inclusion of strong-$CP$ violation in the
baryon sector involves making the same transformations on the baryon
Lagrangian as on the pion Lagrangian, so it is useful to present
a self-contained derivation.  Readers already familiar with the pion $1/\N$
chiral Lagrangian may proceed directly to Sect.~III.

It is well-known that $CP$-violation enters the QCD Lagrangian through
the vacuum angle parameter $\theta$, which is a physical observable of
the theory.  $U(1)_A$ transformations on the QCD Lagrangian can rotate
part or all of this angular dependence amongst the
$G^a_{\mu \nu} \tilde G^{a\, \mu \nu}$ term
and the phase of the quark mass matrix.
Let us adopt the convention
in which all $\theta$-dependence initially resides in the quark mass
matrix\footnote{For intermediate
situations\cite{pich},
$\theta$-dependence appears in the term
$${\theta_0 \over 32 \pi^2 } G^a_{\mu \nu} \tilde G^{a \, \mu \nu}, \quad
\tilde G^{a}_{\mu \nu} \equiv {1 \over 2} \epsilon_{\mu \nu \rho \sigma}
G^{a \, \rho \sigma} $$
in the QCD Lagrangian as well as the phase of the quark mass matrix, such that
$\theta = \theta_0 + {\rm arg}({\rm det} \M_q)$.  Under $U(1)_A$
transformations $R = L^\dagger = e^{i \alpha /2} I$ for $F$ light quark
flavors,
$$\theta_0 \rightarrow \theta_0 + F \alpha,
\quad \M_q \rightarrow e^{-i\alpha} \M_q, $$
leaving $\theta$ invariant.}.
With this convention, the pion nonet chiral Lagrangian is given by
\begin{eqnarray}\label{pione}
{\L}_{\rm pion} &&= {f_\pi^2 \over 4} \left[
\Tr \,\partial_\mu \overline \Sigma \partial^\mu \overline \Sigma^\dagger
- {a \over \N} \left( {i \over 2} \Tr \left( \ln \overline \Sigma
- \ln \overline \Sigma^\dagger \right) \right)^2
\right. \nonumber\\
&&+ \left.
b \Tr \left( \M_q \overline \Sigma + \M_q^\dagger \overline \Sigma^\dagger
\right)
+{c \over \N} \left( \Tr\, \overline \Sigma^\dagger
\partial_\mu \overline \Sigma \right)^2
+ \ldots
\right],
\end{eqnarray}
where $\M_q$ is the quark mass matrix and
$\overline \Sigma = e^{2i\bbox{\Phi}/f_\pi}$ depends nonlinearly on
the pion nonet field $\bbox{\Phi} = \pi^a {\lambda^a \over 2}
+ \eta^\prime {I \over \sqrt{6}}$
divided by $f_\pi = 93$~MeV.  The $\lambda^a$ are the eight
Gell-Mann matrices and $I$ is the $3 \times 3$ unit matrix.  Thus,
the octet component of $\bbox{\Phi}$ is given by
\begin{equation}
{1 \over \sqrt{2}} \left(
\begin{array}{cccc}
{1 \over \sqrt{2}} \pi^0 + {1 \over \sqrt{6}} \eta & \pi^+ & K^+   \\
\pi^- & -{1 \over \sqrt{2}} \pi^0 + {1 \over \sqrt{6}} \eta & K^0  \\
K^- & \bar K^0 & -{2 \over \sqrt{6}} \eta
\end{array}
\right) \quad .
\end{equation}
Under $SU(3)_L \times SU(3)_R$ transformations,
$\overline \Sigma \rightarrow L \overline \Sigma R^\dagger$.
Eq.~(\ref{pione}) is the most general Lagrangian consistent with
chiral and planar QCD flavor symmetry and violation,
to second order in the derivative expansion and to
lowest non-trivial order in $\M_q$ and $1/\N$.  The term proportional
to the parameter $b$ is the usual quark mass term of the pion Lagrangian
which explicitly breaks $SU(3)_L \times SU(3)_R \rightarrow SU(3)$.
The $a$ term, the anomaly term, breaks $U(1)_A$ and is explicitly order
$1/\N$ since the anomaly involves a single quark loop.  The $a$ term
also violates planar QCD flavor symmetry.
The $c$ term splits $f_{\eta^\prime}$ from $f_\pi$.
Thus, the $c$ term violates planar QCD nonet symmetry, and
is explicitly order $1/\N$.  Both the $a$ and $c$ terms preserve the
$U(1)_V$ subgroup of planar QCD flavor symmetry.
The parameters $a$ and $b$ are dimensionful:
$a$ is $\O(\Lambda^2)$ and $b$ is $\O(\Lambda)$, where $\Lambda$ is a
hadronic scale.  The parameter $c$ is dimensionless.
Finally note that
the pion Lagrangian~(\ref{pione}) is of the form
\begin{equation}
\N \ {\L}\left( {\Phi \over \sqrt{\N}}\right) ,
\end{equation}
as required by large-$\N$ power counting rules for mesons,
since $f_\pi \sim \sqrt{\N}$.  It will often be convenient to perform the
rescaling $f_\pi \rightarrow \sqrt{\N} f$ to keep all $\N$-dependence
manifest.

Recall that all $\theta$-dependence of the theory presently resides in the
quark mass matrix $\M_q$, so $\theta = {\rm arg} ({\rm det} \M_q )$.
By performing $SU(3)_L \times SU(3)_R$
transformations, the mass matrix can be written in the form
\begin{equation}
\M_q = \M e^{i\theta/3},
\end{equation}
where $\M = {\rm diag} (m_u, m_d, m_s)$ is real,
since all terms in Lagrangian~(\ref{pione}) are invariant under $SU(3)_L
\times SU(3)_R$ transformations exception for the quark
mass terms which violate chiral symmetry explicitly.  Now remove the phase
in $\M_q$ by performing a $U(1)_A$ transformation,
\begin{equation}
\overline \Sigma \rightarrow e^{-i \theta/3} \overline \Sigma .
\end{equation}
All terms are invariant under this transformation
except for the $a$ term and terms containing the quark mass matrix, so
the Lagrangian becomes
\begin{eqnarray}\label{pitwo}
&&{\L}_{\rm pion} = {f_\pi^2 \over 4} \left[
\Tr \,\partial_\mu \overline \Sigma \partial^\mu \overline \Sigma^\dagger
- {a \over \N} \left( \theta +{i \over 2} \Tr \left( \ln \overline \Sigma
- \ln \overline \Sigma^\dagger \right) \right)^2
\right. \nonumber\\
&&\ + \left.
b\, \Tr \left( \M \left( \overline \Sigma + \overline \Sigma^\dagger \right)
\right)
+ {c \over \N} \left( \Tr\, \overline \Sigma^\dagger \partial_\mu
\overline \Sigma \right)^2
+ \ldots
\right] \ ,
\end{eqnarray}
where the $\theta$-dependence of the Lagrangian is now manifest.
Lagrangian~(\ref{pitwo}) is the chiral Lagrangian with the convention that
no $\theta$-dependence resides in the quark mass matrix.

The $\overline \Sigma$ field spontaneously breaks the $SU(3)$ chiral
symmetry down to its diagonal subgroup.
The vacuum expectation value of $\overline \Sigma$ is determined by
minimization of the potential of the pion Lagrangian.  The potential
$V(\overline \Sigma)$ is given by minus the non-derivative terms in the
Lagrangian.  Since the real mass matrix $\M$ is diagonal, the minimum
of $\overline \Sigma$ is also diagonal, so one looks for a solution of the
form
\begin{equation}
\left< \overline \Sigma \right> = \left(
\begin{array}{cccc}
e^{i\phi_u} & 0 & 0   \\
0 & e^{i\phi_d} & 0  \\
0 & 0 & e^{i\phi_s}
\end{array}
\right) \ .
\end{equation}
The potential as a function of the $\phi_i$ is
\begin{equation}
V(\phi_i) = {f_\pi^2 \over 4} \left( - \sum_i 2b \, m_i \, {\rm cos}\phi_i
+ {a \over \N} \left( \theta - \sum_i \phi_i \right)^2 \right) \ .
\end{equation}
Minimization of the potential leads to the equations
\begin{equation}\label{minphi}
2 b \, m_i \, {\rm sin} \phi_i
= {a \over \N} \left( \theta - \sum_{j} \phi_j \right) .
\end{equation}
The solution of Eq.~(\ref{minphi}) for a given $\theta$ determines the
angles $\phi_i$ as a function of $\theta$.  Solutions of
Eq.~(\ref{minphi}) for different values of $\theta$ are discussed in
detail in Refs.~\cite{dVV,witten}.

It is more physical to reexpress the Lagrangian in terms of a
$\Sigma$ field with vacuum expectation value $\left< \Sigma \right> = I$;
this vacuum realignment is performed by making the substitution
\begin{equation}\label{vacalign}
\overline \Sigma = \left< \overline \Sigma \right> \Sigma
\end{equation}
in Eq.~(\ref{pitwo}), so that
\begin{eqnarray}\label{pithree}
{\L}_{\rm pion} &&= {f_\pi^2 \over 4} \left[
\Tr \,\partial_\mu \Sigma \partial^\mu \Sigma^\dagger
- {a \over \N} \left( \overline \theta +{i \over 2} \Tr \left( \ln \Sigma
- \ln \Sigma^\dagger \right) \right)^2
\right. \nonumber\\
&&+ \left.
b \,\Tr \left( \overline \M \Sigma + \overline \M^\dagger \Sigma^\dagger
\right)
+ {c \over \N} \left( \Tr\, \Sigma^\dagger \partial_\mu \Sigma \right)^2
+ \ldots
\right] \ ,
\end{eqnarray}
where
\begin{equation}\label{massalign}
\overline \M = \M \left< \overline \Sigma \right>
= {\rm diag} \left( m_i \, e^{\phi_i} \right) \ ,
\end{equation}
and
\begin{equation}
\overline \theta = \left( \theta - \sum_i \phi_i \right) \ .
\end{equation}
Using the minimization equation~(\ref{minphi}), the mass can be
rewritten as
\begin{equation}\label{mrealim}
\overline \M = \M(\theta)
+ i \,{ {a \overline \theta} \over {2b}} {1 \over \N} I,
\end{equation}
where
\begin{equation}
\M(\theta) = {\rm diag} \left( m_i \, {\rm cos} \phi_i \right) \ .
\end{equation}
Using Eq.~(\ref{mrealim}), one obtains the final version of the
pion Lagrangian,
\begin{eqnarray}\label{pifour}
{\L}_{\rm pion} &&= {f_\pi^2 \over 4} \left[
\Tr \,\partial_\mu \Sigma \partial^\mu \Sigma^\dagger
- {a \over \N} \left( {i \over 2} \Tr \left( \ln \Sigma
- \ln \Sigma^\dagger \right) \right)^2
\right. \nonumber\\
&&+
b \,\Tr \left( \M(\theta) \left( \Sigma + \Sigma^\dagger - 2 \right)
\right)
+ {c \over \N} \left( \Tr\, \Sigma^\dagger \partial_\mu
 \Sigma \right)^2 \\
&&+ \left.
i { {a \overline \theta} \over \N} \left( {1 \over 2} \Tr \left( \Sigma
- \Sigma^\dagger \right) - \Tr \left( \ln \Sigma - \ln \Sigma^\dagger \right)
\right) + \ldots
\right] \ ,\nonumber
\end{eqnarray}
where a constant term has been dropped relative to Eq.~(\ref{pithree}).

The observed spectrum and mixing of the pion nonet can be understood using
Lagrangian~(\ref{pifour}) if the parameters satisfy \cite{dVV}
\begin{equation}
b m_u, \ b m_d \ll b m_s < {a \over \N} \ .
\end{equation}
In the $\bar u u$, $\bar d d$ and $\bar s s$ basis, the neutral meson mass
matrix is given by
\begin{equation}\label{pimass}
b \left(
\begin{array}{cccc}
m_u {\rm cos}\phi_u & 0 & 0   \\
0 & m_d {\rm cos}\phi_d & 0  \\
0 & 0 & m_s {\rm cos}\phi_s
\end{array}
\right)
+ {a \over \N} \left(
\begin{array}{cccc}
1 & 1 & 1   \\
1 & 1 & 1  \\
1 & 1 & 1
\end{array}
\right) ,
\end{equation}
to leading order in explicit chiral symmetry breaking and
$1/\N$\footnote{The effects of higher order terms on a leading order
bound on the mass ratio
$\eta/\eta^\prime$\cite{hgeorgi} have been considered
recently in Ref.~\cite{peris}.}.
In the chiral limit $m_i \rightarrow 0$, the $\eta^\prime$ is a massive
$SU(3)$ singlet with mass $m_{\eta^\prime}^2 = F a/\N$, and
the octet mesons are exact massless Goldstone bosons\cite{dVV}.

\section{Baryon Chiral Lagrangian}
This section formulates a $1/\N$ baryon chiral Lagrangian for
$\N$ large, finite and odd.  The Lagrangian is first presented in the flavor
symmetry limit.  Explicit flavor symmetry breaking terms involving the quark
mass matrix are then added to the Lagrangian.  Strong-CP violation
enters the baryon chiral Lagrangian through these terms.  The baryon chiral
Lagrangian is written in terms of the $1/\N$ (spin $\otimes$ flavor) operator
expansion for baryons.  The structure of this operator expansion is
reviewed below.

The (spin $\otimes$ flavor) $1/\N$ expansion for baryons
organizes the lowest-lying baryon states into the completely symmetric
$SU(2F)$ representation shown in Fig.~1.  Under  $SU(2) \otimes SU(F)$
symmetry, this representation decomposes into a tower of baryon states
with spins $1/2, \ldots, \N/2$ in the flavor representations displayed in
Fig.~2.  The weight diagrams of the flavor representations of the
spin-$1/2$ and spin-$3/2$ baryons for $F =3$ are given in Figs.~3 and~4,
respectively.  For $\N=3$, these flavor multiplets reduce to the baryon octet
and decuplet, but for $\N>3$, the multiplets contain additional baryon states
which do not exist for $\N=3$.  Because of the complexity of the flavor
representations for $F > 2$, it is easier to focus on the operators than
the states.

Any QCD operator transforming according to a given
$SU(2) \times SU(F)$ representation has an expansion in terms of
$n$-body operators of the form,
\begin{equation}\label{qcdop}
{\cal O}_{QCD}= \sum_{n} c_{(n)} {1\over \N^{n-1}} {\cal O}_n ,
\end{equation}
where the operator basis ${\cal O}_n$ consists of polynomials in
the spin-flavor generators $J^i$, $T^a$, and $G^{ia}$.  The operator
coefficients $c_{(n)}(1/\N)$ have power series expansions in $1/\N$
beginning at order unity.

The problem of finding a complete and independent set of operators
for any spin-flavor representation was solved in Ref.~\cite{djmtwo}.
The basic building blocks of the expansion are the $0$-body
$SU(2F)$ identity operator $\openone$ and the $1$-body operators
$J^i$, $T^a$ and $G^{ia}$
which satisfy the $SU(2F)$ commutation relations.  Because antisymmetric
products of these operators
can be reduced using the commutation relations, one only needs
to consider operator products which are completely symmetric in
non-commuting operators.  In addition, it suffices to keep polynomials
through order $\N$ for the lowest-lying baryons.  There are
a number of identities among the
polynomials of order less than or equal to $\N$ which further reduce the
operator basis.  The complete set of identities were derived in
Ref.~\cite{djmtwo} using quark operators
\begin{eqnarray}
J^i & = &
q^\dagger \left( {\sigma^i \over 2} \otimes I \right) q\qquad (1,1),
\nonumber \\
T^a & = &
q^\dagger \left(I \otimes {\lambda^a \over 2}\right) q\qquad
(0,8),\\
G^{ia} & = &
q^\dagger \left({\sigma^i \over 2} \otimes {\lambda^a \over 2}\right) q
\qquad (1, 8)\ .
\nonumber
\end{eqnarray}
This work also uses the quark representation of the $1/\N$ operator expansion
for baryons.  Equivalent results can be obtained in the Skyrme representation.

\subsection{Lagrangian in the Flavor Symmetry Limit}

In the large-$\N$ limit, baryons have masses of order $\N$ and become
very heavy relative to mesons with masses of order $1$.
The $1/\N$ baryon chiral Lagrangian is formulated treating baryons as
heavy static fields with fixed velocity $v^\mu$~\cite{bcpt,eh,georgi}.
The $1/\N$ expansion provides a systematic expansion parameter for this
procedure.  The following $1/\N$
chiral Lagrangian is written in the rest frame of the baryon, which
is the natural frame for the
(spin $\otimes$ flavor) operator expansion.  The generalization to an arbitrary
velocity frame is straightforward.

 The $1/\N$ baryon chiral Lagrangian for arbitrary $\N$ is of the form
\begin{eqnarray}\label{bary}
\L_{\rm baryon} &&= i \D^0 - M_{\rm hyperfine}
+ \Tr \left( \A^i \lambda^a \right) A^{ia} \nonumber\\
&&\quad\quad + \Tr \left( \A^i {{2I} \over \sqrt{6}} \right) A^{i} + \ldots,
\end{eqnarray}
with
\begin{equation}\label{cov}
\D^0= \partial^0 \openone + \Tr \left(\V^0 \lambda^a \right) T^a
+{1 \over 3} \Tr \left( \V^0 I \right) \N \openone .
\end{equation}

The notation of Eqs.~(\ref{bary}) and~(\ref{cov}) is very compact:
each term involves
a baryon operator.  The baryon kinetic energy term is proportional to the
spin-flavor identity element $\openone$.  The hyperfine baryon mass operator
describes the spin splittings of the baryon tower.
Pion fields appear in the chiral Lagrangian through
the vector and axial vector combinations
\begin{eqnarray}\label{picurrents}
\V^0 &=& {1 \over 2} \left( \xi \partial^0 \xi^\dagger
+ \xi^\dagger \partial^0 \xi \right),\nonumber\\
\A^i &=& {i \over 2} \left( \xi \grad i \xi^\dagger
- \xi^\dagger \grad i\xi \right),
\end{eqnarray}
which depend nonlinearly on the field $\xi =e^{i \bbox{\Phi} / f_\pi}$.
The vector pion combinations couple to baryon vector charges:
the flavor octet pion combination couples to the flavor octet baryon
charge\footnote{The subscript QCD is used to emphasize that
the quark fields are QCD quark fields, not
the quark creation and annihilation operators of the quark representation.}
\begin{equation}
V^{0a} = \left< \B^\prime \left|
\left( \overline q \gamma^0 {\lambda^a \over 2} q \right)_{\rm QCD}
\right|\B \right>,
\end{equation}
while the flavor singlet pion combination
\begin{equation}
\Tr \left( \V^0 {{2I} \over \sqrt{6}} \right)
\end{equation}
couples to the flavor singlet baryon charge
\begin{equation}
V^{0} = \left< \B^\prime \left|
\left( \overline q \gamma^0 {I \over \sqrt{6} } q \right)_{\rm QCD}
\right|\B \right>.
\end{equation}
The baryon vector charges equal
\begin{eqnarray}
V^{0a} &&= v^0 T^a = T^a, \nonumber\\
V^0 &&= v^0 {1 \over \sqrt{6}} \N \openone
= {1 \over \sqrt{6}} \N \openone,
\end{eqnarray}
to all orders in the $1/\N$ expansion.
The $\ell=1$ flavor octet axial vector pion combination couples to the flavor
octet baryon axial current
\begin{equation}\label{axi}
A^{ia} = \left< \B^\prime \left|
\left( \overline q \gamma^i \gamma_5 {\lambda^a \over 2} q \right)_{\rm QCD}
\right|\B \right>,
\end{equation}
whereas the flavor singlet axial
pion combination couples to the flavor singlet baryon
axial current
\begin{equation}
A^{i} = \left< \B^\prime \left|
\left( \overline q \gamma^i \gamma_5 {I \over \sqrt{6} }q \right)_{\rm QCD}
\right|\B \right> .
\end{equation}
The ellipses in Eq.~(\ref{bary}) denotes higher partial wave pion
couplings which occur at subleading orders in the $1/\N$ expansion for $\N>3$.
At leading order in the $1/\N$ expansion, the pion couplings of baryons
are purely $p$-wave for any $\N$ \cite{djmtwo}.

The baryon chiral Lagrangian describes the interactions of the pions and
baryons in terms of QCD baryon operators.  Each of these operators has
an expansion in $1/\N$ of the form Eq.~(\ref{qcdop}).

In the limit of exact $SU(3)$ flavor symmetry,
the baryon mass operator is defined by
\begin{equation}
M = \left< \B^\prime \left|
\H_{\rm QCD}
\right|\B \right>,
\end{equation}
where $\H_{\rm QCD}$ is the QCD Hamiltonian in the chiral limit
$m_i \rightarrow 0$.  The baryon mass operator transforms as a $(0,1)$
under $SU(2) \times SU(3)$ symmetry.  The $1/\N$ expansion for a
$(0,1)$ QCD operator is of the form\cite{j,djm,cgo,lm}
\begin{equation}\label{massop}
M = m^{0,1}_{(0)} \N \openone
+ \sum_{n=2,4}^{\N-1} m^{0,1}_{(n)} {1 \over \N^{n-1}} J^n.
\end{equation}
The coefficients $m^{0,1}_{(n)}$ are dimensionful parameters
of ${\cal O}(\Lambda)$.
The first term in expansion~(\ref{massop}),
the overall spin-independent mass of the
baryon multiplet, is removed from the chiral Lagrangian by the heavy baryon
field
redefinition \cite{bcpt}.
The spin-dependent terms in Eq.~(\ref{massop}) define $M_{\rm hyperfine}$,
which
appears explicitly in the Lagrangian.  The hyperfine mass expansion
reduces to a single operator \cite{j}
\begin{equation}
M = m^{0,1}_{(2)} {1 \over \N} J^2 \ ,
\end{equation}
for $\N=3$.

The $1/\N$ expansions for the baryon flavor octet and singlet axial currents
were derived in Ref.~\cite{djmtwo}.  The $1/\N$ expansion for the
$(1,8)$ baryon axial current is given by
\begin{equation}\label{axin}
A^{ia} = a^{1,8}_{(1)} G^{ia} + \sum_{n=2,3}^{\N} b^{1,8}_{(n)}
{1 \over \N^{n-1} } {\cal D}_n^{ia}
+ \sum_{n=3,5}^{\N} c^{1,8}_{(n)} {1 \over \N^{n-1} } {\cal O}_n^{ia},
\end{equation}
where the ${\cal D}_n^{ia}$ are diagonal operators, with nonzero matrix
elements only between states with the same spin, and the ${\cal O}_n^{ia}$
are purely off-diagonal operators, with nonzero matrix elements only between
states with different spin.  The operators ${\cal D}_n^{ia}$ and
${\cal O}_n^{ia}$ are defined in Ref.~\cite{djmtwo}.
Eq.~(\ref{axin}) reduces to
\begin{eqnarray}\label{aia}
A^{ia} &=& a^{1,8}_{(1)} G^{ia}
+ b^{1,8}_{(2)} {1 \over \N} J^i T^a
+ b^{1,8}_{(3)} {1 \over \N^2} \{ J^i \{ J^j, G^{ja} \} \}\nonumber\\
&&+ c^{1,8}_{(3)} {1 \over \N^2} \left(
\{J^2, G^{ia}\} - {1 \over 2} \{ J^i \{ J^j, G^{ja} \} \}
\right)
\end{eqnarray}
for $\N=3$.
The $1/\N$ expansion for the $(1,1)$ baryon axial current is given by
\begin{equation}\label{saxin}
A^{i} = \sum_{n=1,3}^{\N} b^{1,1}_{(n)} {1 \over \N^{n-1}} {\cal D}^i_n ,
\end{equation}
where ${\cal D}^i_1 = J^i$ and ${\cal D}^i_{n+2} = \{ J^2, {\cal D}^i_n \}$.
Eq.~(\ref{saxin}) reduces to
\begin{equation}
A^{i} = b^{1,1}_{(1)} J^i + b^{1,1}_{(3)} {1 \over \N^2}\{J^2, J^i \} \ .
\end{equation}
for $\N=3$.

Lagrangian~$\L_{\rm baryon}$ is the most general Lagrangian
invariant under $SU(3)_L \times SU(3)_R \times U(1)_V \times U(1)_A$
chiral symmetry and contracted spin-flavor symmetry.  The form of
the Lagrangian
factorizes baryon invariants from pion invariants explicitly, which is
necessary because baryons transform under a larger symmetry than mesons
in the large-$\N$ limit.  The Lagrangian correctly
relates baryon-multipion vertices using chiral symmetry.
Under chiral
transformations,
\begin{equation}\label{xitrans}
\xi \rightarrow L \xi U^\dagger = U \xi R^\dagger,
\end{equation}
where $U$ is a vector $SU(3)\times U(1)$ transformation
defined by Eq.~(\ref{xitrans}).
The flavor representations of the baryon spin tower transform as $\N$-index
tensor representations under $U$,
\begin{equation}
B^{\alpha_1\ldots\alpha_n\ldots\alpha_{\N}}
\rightarrow \sum_{n} U^{\alpha_n}_{\alpha_n^\prime}
B^{\alpha_1\ldots\alpha_n^\prime\ldots\alpha_{\N}},
\end{equation}
where the symmetry of the baryon flavor tensors is dictated by the
Young tableaux of Fig.~2.
The pion combinations~(\ref{picurrents}) are
unaffected by vacuum realignment~(\ref{vacalign}), so the $SU(3)$-symmetric
baryon chiral Lagrangian contains no strong $CP$ violation.

Planar QCD
flavor symmetry further constrains the parameters of the
$1/\N$ baryon chiral Lagrangian.  In the next subsection,
planar QCD flavor symmetry is imposed
on the baryon chiral Lagrangian.  In the following
subsection, the chiral Lagrangian for the baryon octet and decuplet
is compared with the $1/\N$ baryon chiral Lagrangian at $\N=3$.

\subsubsection{Planar QCD Flavor Symmetry}

Planar QCD flavor symmetry implies that the baryon $1/\N$ chiral Lagrangian
possesses a $SU(2) \otimes U(3)$ (spin $\otimes$ flavor) symmetry
at leading order in the $1/\N$ expansion.  The symmetry is broken at
first subleading order by diagrams with a single quark loop, as shown in
Fig.~5.

Planar QCD flavor symmetry constrains $\L_{\rm baryon}$ by forming a
nonet baryon axial vector current out of the singlet and octet baryon
axial vector currents at leading order in the $1/\N$
expansion\footnote{The baryon vector currents $V^{0a}$ and $V^0$ form
a flavor nonet to all orders in the $1/\N$ expansion.},
\begin{equation}\label{nonetaxi}
A^i = A^{i9} + O(1/\N) .
\end{equation}
This constraint relates the coefficients of the $A^i$ expansion
to those of the $A^{ia}$ expansion in the limit $\N \rightarrow \infty$.
The easiest way to impose Eq.~(\ref{nonetaxi}) is to replace the
operator coefficients of the singlet axial vector expansion Eq.~(\ref{saxin})
by
\begin{equation}
b^{1,1}_{(n)} \rightarrow \overline b^{1,1}_{(n)}
+ {1 \over \N}b^{1,1}_{(n)},
\end{equation}
where the coefficients with an overscore are determined by exact nonet
symmetry, and the remainders are unconstrained and violate nonet symmetry
at first subleading order $1/\N$.
For arbitrary $\N$, nonet symmetry implies
\begin{eqnarray}
\overline b^{1,1}_{(1)} &=& {1 \over \sqrt{6}} \left( a^{1,8}_{(1)}
+ b^{1,8}_{(2)}\right), \nonumber\\
\overline b^{1,1}_{(3)} &=& {1 \over \sqrt{6}} \left( 2 b^{1,8}_{(3)}
+ b^{1,8}_{(4)} \right), \nonumber\\
\overline b^{1,1}_{(5)} &=& {1 \over \sqrt{6}} \left( 2 b^{1,8}_{(5)}
+ b^{1,8}_{(6)} \right), \\
&\ldots&\ldots  \nonumber
\end{eqnarray}
etc., where the relative factor of $1/\sqrt{6}$ occurs because the ninth
flavor components of $G^{ia}$ and $T^a$ are related to $J^i$ and $\N \openone$
by
\begin{eqnarray}\label{gtnine}
G^{i9} & = &
q^\dagger \left({\sigma^i \over 2} \otimes {I \over \sqrt{6}}\right) q
= {1 \over \sqrt{6}} J^i \qquad (1, 1), \nonumber \\
T^9 & = &
q^\dagger \left(I \otimes {I \over \sqrt{6} }\right) q
= {1 \over \sqrt{6}} \N \openone \qquad
(0,1).
\end{eqnarray}
Notice that the coefficients of the diagonal operators ${\cal D}_n^i$
in the singlet expansion do not depend on the coefficients $c^{1,8}_{(n)}$
of the off-diagonal
operators ${\cal O}_n^{ia}$ in the octet expansion.
For $\N =3$, the nonet symmetry conditions reduce to
\begin{eqnarray}\label{axiconstraint}
\overline b^{1,1}_{(1)} &=& {1 \over \sqrt{6}} \left( a^{1,8}_{(1)}
+ b^{1,8}_{(2)}\right), \nonumber\\
\overline b^{1,1}_{(3)} &=& {1 \over \sqrt{6}} \left( 2 b^{1,8}_{(3)}\right),
\end{eqnarray}
where the second condition is modified because the $4$-body operator
corresponding to $b^{1,8}_{(4)}$ does not occur in the operator basis for
$\N=3$.

It is important to stress that the nonet symmetry
constraint Eq.~(\ref{nonetaxi}) leads to a condition for each operator
coefficient in the singlet expansion since this constraint must be
satisfied for {\it all} spin states of the baryon tower
(not just the states with spins of order unity).
The fact that Eq.~(\ref{nonetaxi}) is satisfied operator by operator
in the baryon (spin $\otimes$ flavor) operator expansion is consistent
with the violation of planar QCD flavor symmetry by single quark
loop diagrams Fig.~5, since this breaking is decoupled from
the baryon $1/\N$ operator expansion.

The final version of the $1/\N$ baryon chiral Lagrangian can be
obtained by rewriting $\L_{\rm baryon}$ is a form which
implements the constraints of planar QCD symmetry explicitly,
\begin{eqnarray}\label{barytwo}
\L_{\rm baryon} &&= i \D^0 - M_{\rm hyperfine}
+ \Tr \left( \A^i \lambda^a \right) A^{ia} \nonumber\\
&&\quad\quad + {1 \over \N} \Tr \left( \A^i {{2I} \over \sqrt{6}} \right)
A^{i} + \ldots,
\end{eqnarray}
with
\begin{equation}\label{covtwo}
\D^0= \partial^0 \openone + \Tr \left(\V^0 \lambda^a \right) T^a ,
\end{equation}
where $a=1, \ldots,9$, $\lambda^9 \equiv 2 I/ \sqrt{6}$, and the
baryon $1$-body operators $T^9$ and $G^{i9}$ are defined in
Eq.~(\ref{gtnine}).  Nonet flavor symmetry of
the baryon-pion axial vector couplings is
broken by the last term, which gives a nonet symmetry-breaking
contribution to the singlet current at relative order $1/\N$.

\subsubsection{Comparison with Octet and Decuplet Chiral Lagrangian}

It is instructive to compare the $1/\N$ chiral Lagrangian
at $\N=3$ with the chiral Lagrangian for the baryon octet and decuplet
without a $1/\N$ expansion.  The flavor octet pion couplings of
the octet and decuplet are described by the chiral Lagrangian\cite{bcpt},
\begin{eqnarray}\label{chibary}
{\L}_{\rm baryon} & = & i \Tr \overline B_v \left( v \cdot \D\right) B_v
-i \overline T_v^\mu \left( v \cdot \D\right) T_{v\,\mu}
+ \Delta \overline T_v^\mu T_{v\,\mu} \nonumber\\
&&+ 2 D\, \Tr \overline B_v \S_v^\mu \left\{ \A_\mu, B_v \right\}
+ 2 F\, \Tr \overline B_v \S_v^\mu \left[ \A_\mu, B_v \right] \nonumber\\
&&+ \C\, \left( \overline T_v^\mu \A_\mu B_v + \bar B_v \A_\mu T_v^\mu \right)
\\
&&+ 2 \H\, \overline T_v^\mu \S_{v}^{\nu} \A_\nu T_{v\,\mu},\nonumber
\end{eqnarray}
where $D$, $F$, ${\cal C}$ and ${\cal H}$ are the baryon-pion couplings and
$\Delta = m_T - m_B$ is the decuplet-octet mass difference.
The octet mass $m_B$ has been removed from
the Lagrangian by the heavy baryon field redefinition.
Flavor singlet baryon-$\eta^\prime$ couplings can be incorporated
into the chiral Lagrangian by adding two terms,
\begin{eqnarray}\label{chis}
2 S_B\, \Tr \A_\mu \, \Tr \overline B_v \S_v^\mu B_v
- 2 S_T\, \Tr \A_\nu \, \overline T_v^\mu \S_{v}^{\nu} T_{v\,\mu}\ ,
\end{eqnarray}
where $S_B$ and $S_T$ are the singlet axial coupling constants of the
octet and decuplet, respectively.

There is a one-to-one correspondence between the parameters of the
octet and decuplet chiral Lagrangian and the coefficients of
the $1/\N$ baryon chiral Lagrangian at $\N=3$.  The mass parameters
are related to the $1/\N$ mass coefficients by
\begin{eqnarray}\label{mbmt}
m_B &&= 3 m^{0,1}_{(0)} + {1 \over 4} m^{0,1}_{(2)}, \nonumber\\
m_T &&= 3 m^{0,1}_{(0)} + {5 \over 4} m^{0,1}_{(2)},
\end{eqnarray}
so that
\begin{equation}
\Delta = m^{0,1}_{(2)}\ .
\end{equation}
  The flavor octet baryon-pion
couplings are related to the coefficients of the $1/\N$ expansion at $\N=3$
by
\begin{eqnarray}\label{dfch}
D &=& {1 \over 2} a^{1,8}_{(1)}
+ {1 \over {6}} b^{1,8}_{(3)}, \nonumber\\
F&=& {1 \over 3} a^{1,8}_{(1)}
+ {1 \over 6} b^{1,8}_{(2)}
+ {1 \over 9} b^{1,8}_{(3)}, \nonumber\\
{\cal C} &=& -a^{1,8}_{(1)}
- {1 \over 2} c^{1,8}_{(3)},\\
{\cal H} &=& -{3 \over 2} a^{1,8}_{(1)}
-{3 \over 2} b^{1,8}_{(2)}
- {5 \over 2} b^{1,8}_{(3)}. \nonumber
\end{eqnarray}
Notice that the purely off-diagonal operator coefficient $c^{1,8}_{(3)}$
contributes only to the octet-decuplet-pion coupling constant ${\cal C}$,
and that the diagonal operator coefficients $b^{1,8}_{(n)}$ contribute
only to the diagonal couplings $D$, $F$, and ${\cal H}$.  In addition,
$b^{1,8}_{(2)}$ is pure $F$, and does not contribute to $D$.
The flavor singlet baryon-pion couplings are related to the
coefficients of the $1/\N$ expansion at $\N=3$ by
\begin{eqnarray}\label{sbst}
S_B &=& {1 \over \sqrt{6}} \left( b^{1,1}_{(1)} + {1 \over 6} b^{1,1}_{(3)}
\right), \nonumber\\
S_T &=& {3 \over \sqrt{6}} \left( b^{1,1}_{(1)} + {5 \over 6} b^{1,1}_{(3)}
\right).
\end{eqnarray}
The factor of $3$ in the second relation occurs because the
decuplet spin operator in Eq.~(\ref{chis}) acts only on the spinor
portion of the spin-$3/2$ Rarita-Schwinger field \cite{bcpt}.  The
metric of the spin-one portion of the spin-$3/2$ field $T_v^\mu$
cancels the minus sign of the decuplet term in Eq.~(\ref{chis}).

Relations~(\ref{mbmt})--(\ref{sbst}) are valid for $\N$ set equal to three.
For arbitrary $\N$, the $1/\N$ expansions for baryons with spins of order unity
can be truncated,
\begin{eqnarray}\label{trunc}
M &&= m^{0,1}_{(0)} \N \openone, \nonumber\\
A^{ia} &&= a^{1,8}_{(1)} G^{ia}
+ b^{1,8}_{(2)} {1 \over \N} J^i T^a, \\
A^i &&= b^{1,1}_{(1)} J^i, \nonumber
\end{eqnarray}
where Eqs.~(\ref{trunc}) are valid up to terms of relative order $O(1/\N^2)$
{\it everywhere} in the flavor weight diagrams.
The parameter $b^{1,8}_{(2)}$
produces deviations from $SU(6)$ symmetry.  In the limit $\N \rightarrow 3$,
Eqs.~(\ref{trunc}) lead to the parameter relations
\begin{eqnarray}
m_B &&= m_T, \nonumber\\
{\cal C} = -2D , &&\quad {\cal H} = 3D - 9F, \\
S_B &&= {1 \over 3} S_T . \nonumber
\end{eqnarray}

The implementation of flavor nonet symmetry on the axial vector
baryon-pion couplings raises an interesting subtlety.
The spin-$1/2$ baryon $SU(3)$
field with mixed symmetry is written as a tensor with an upper index
and a lower index by using the flavor $SU(3)$ $\epsilon$-tensor to replace
two antisymmetric upper indices by a single lower index,
\begin{equation}
B^{\alpha}_{\beta} = \epsilon_{\beta \gamma \delta}
B^{\alpha \left[ \gamma \delta \right]}.
\end{equation}
The octet tensor $B^{\alpha}_{\beta}$
transforms in the same manner as the three-index tensor
$B^{\alpha \left[ \gamma \delta \right]}$ under $SU(3)$
transformations, since the $\epsilon$-tensor is an invariant tensor under
$SU(3)$ transformations.
The $\epsilon$-tensor, however, is not invariant
under $U(1)$ transformations, so replacing
$B^{\alpha \left[ \gamma \delta \right]}$ by $B^{\alpha}_{\beta}$
is not legitimate when $U(3)$ flavor symmetry is present.
Nonet flavor symmetry cannot be imposed on the baryon chiral
Lagrangian~(\ref{chibary}) by simply
promoting $\V_\mu$ and $\A_\mu$ to a nonet matrices since the Lagrangian
is written in terms of $B^{\alpha}_{\beta}$.  It is not difficult
to work out the condition of nonet symmetry for the baryon axial couplings
using $B^{\alpha \left[ \gamma \delta \right]}$,
\begin{eqnarray}\label{snonet}
S_B &&\rightarrow {1 \over 3} \left( 3 F - D \right), \nonumber\\
S_T &&\rightarrow -{1 \over 3} \H \ .
\end{eqnarray}
The consistency of Eq.~(\ref{snonet}) with Eq.~(\ref{axiconstraint})
can be checked using Eqs.~(\ref{dfch}) and~(\ref{sbst}).

The generalization of Eq.~(\ref{snonet}) to arbitrary $\N$ for the
spin-$1/2$ baryons also is of interest.
Ref.~\cite{djm} defined the pion octet couplings ${\cal M}$ and
${\cal N}$ of the spin-$1/2$ baryons in terms of the large-$\N$
baryon tensor with one upper index and $\nu = (\N -1)/2$ lower indices.
The lesson of $U(3)$ symmetry is that
the use of baryon tensors with antisymmetric indices lowered by
the flavor $\epsilon$-symbol is to be avoided in the $1/\N$ expansion;
for general $\N$, one should work exclusively with $\N$-index baryon
flavor tensors.  It is straightforward to rewrite
these invariants using the $\N$-index flavor tensor
\begin{equation}\label{btensor}
B^{\alpha_1 \left[\alpha_2 \alpha_3 \right] \ldots
\left[ \alpha_{\N-1} \alpha_{\N} \right]}
\end{equation}
for the spin-$1/2$
baryons.  Planar QCD flavor symmetry relates the singlet invariant
of the spin-$1/2$ baryons to the octet invariants,
\begin{equation}\label{shalf}
{\cal S}_{1/2} = {1 \over 3} \left( {\cal M} - 2 {\cal N} \right)
+ O\left( {1 \over \N} \right),
\end{equation}
where ${\cal S}_{1/2}$ is the generalization of $S_B$ for large-$\N$
flavor representations.  The $O(1/\N)$ correction to Eq.~(\ref{shalf})
is due to violation of planar QCD flavor symmetry\footnote{The invariant
${\cal S}_{1/2}$ is $O(1)$ even though ${\cal M}$ and ${\cal N}$ are
both $O(\N)$, so the correction to Eq.~(\ref{shalf}) is both
of relative and absolute order $1/\N$.}.
Ref.~\cite{djm} proved that the ratio
\begin{equation}
{ {\cal N} \over {\cal M}} = {1 \over 2} + {\alpha \over \N}
+ O\left({1 \over {\N^2}} \right),
\end{equation}
where ${\cal M}$ and ${\cal N}$ are both $O(\N)$.
Substitution into Eq.~(\ref{shalf}) shows that the
leading term cancels, so that
\begin{equation}\label{shalftwo}
{ {\cal S}_{1/2} \over {\cal M} } = - {2 \over 3} {\alpha \over \N}
+ O\left({1 \over {\N^2}} \right) ,
\end{equation}
where the $O(1/\N^2)$ correction depends on nonet symmetry violation and
on the $O(1/\N^2)$ contribution to ${\cal N}/ {\cal M}$.
Ref.~\cite{djmtwo} showed that
\begin{equation}
\alpha = - {3 \over 2} \left( 1 +
{ {b^{1,8}_{(2)}} \over {a^{1,8}_{(1)}} } \right),
\end{equation}
so Eq.~(\ref{shalftwo}) implies that
the singlet axial current is order $1/\N$ relative to the octet current
and that the normalization depends on the ratio
of $b_{(2)}^{1,8}$ to $a_{(1)}^{1,8}$ at leading order.

\subsection{Lagrangian with Quark Mass Flavor Breaking}

Explicit flavor symmetry breaking enters the
baryon chiral Lagrangian through terms containing powers of the
quark mass matrix.  The leading Lagrangian with a
single insertion of the quark mass matrix is presented
in this subsection.
The singlet and octet components of
these linear terms form a nonet at leading order in the $1/\N$
expansion due to planar QCD flavor symmetry.  Vacuum realignment
generates strong-$CP$ violating terms, which also form a nonet
at leading order in the $1/\N$ expansion.
Subsection B1 compares
the $1/\N$ Lagrangian terms with one insertion of the quark mass
matrix to the octet and decuplet Lagrangian with no $1/\N$ expansion.
Subsection B2 discusses the implications of nonet symmetry for
the proton matrix
element $\left< p \left| m_s \overline s s \right| p \right>$.

The leading Lagrangian with one power of the quark mass matrix is
given by
\begin{eqnarray}\label{masslag}
\L^{\M}_{\rm baryon} &&=
\Tr \left( \left( \M \overline \Sigma
+ \M^\dagger \overline \Sigma^\dagger \right)
{I \over \sqrt{6}} \right) \H^0 \nonumber\\
&& +\Tr \left( \left( \overline \xi \M \overline \xi
+ \overline \xi^\dagger \M^\dagger \overline \xi^\dagger \right)
{\lambda^3 \over 2} \right)
\H^3 \nonumber\\
&& +\Tr \left( \left( \overline \xi \M \overline \xi
+ \overline \xi^\dagger \M^\dagger \overline \xi^\dagger \right)
{\lambda^8 \over 2} \right)
\H^8 \ ,
\end{eqnarray}
where the singlet perturbation to the Hamiltonian
\begin{equation}
\H^0 = {1 \over \sqrt{6}} \left< \B^\prime \left|
\left( \overline q q \right)_{\rm QCD}
\right|\B \right>,
\end{equation}
and the octet $a=3,8$ Hamiltonian perturbations
\begin{equation}
\H^{a} = \left< \B^\prime \left|
\left( \overline q {\lambda^a \over 2} q \right)_{\rm QCD}
\right|\B \right>\ .
\end{equation}
Note that terms containing the pseudocalar mass combination
$(\overline \xi \M \overline \xi
- \overline \xi^\dagger \M^\dagger \overline \xi^\dagger )$
are subleading in the $1/\N$ expansion and have been neglected.
These terms
are suppressed by one factor of $1/\N$ relative to the terms
involving $(\overline \xi \M \overline \xi
+ \overline \xi^\dagger \M^\dagger \overline \xi^\dagger )$
since baryon matrix elements of the pseudoscalar QCD
quark operators are $O(1/\N)$.

The explicit symmetry breaking perturbations to the baryon
Hamiltonian have expansions in $1/\N$.  The general expansion of the singlet
perturbation has the same form as Eq.~(\ref{massop}), and reduces to
\begin{equation}
\H^0 = b^{0,1}_{(0)} \N \openone + b^{0,1}_{(2)} {1 \over \N} J^2 \nonumber\\
\end{equation}
for $\N=3$.  The general expansion for the $(0,8)$ perturbation was derived
in Ref.~\cite{djmtwo},
\begin{equation}\label{hamop}
\H^a = \sum_{n=1}^{\N} b_{(n)}^{0,8} {1 \over \N^{n-1}} {\cal D}_n^a,
\end{equation}
where ${\cal D}_1^a = T^a$, ${\cal D}_2^a = \{ J^i, G^{ia} \}$ and
${\cal D}_{n+2}^a = \{ J^2, {\cal D}_n^a \}$.  Eq.~(\ref{hamop})
reduces to
\begin{equation}
\H^a = b^{0,8}_{(1)} T^a + b^{0,8}_{(2)} {1 \over \N} \{ J^i, G^{ia} \} +
b^{0,8}_{(3)} {1 \over \N^2} \{ J^2, T^a \}
\end{equation}
for $\N=3$.

Vacuum realignment affects the
quark mass terms, resulting in
baryon-pion couplings which violate strong $CP$.
Eqs.~(\ref{vacalign}) and~(\ref{massalign}) imply that
the mass combination appearing in Eq.~(\ref{masslag}) is replaced by\cite{pich}
\begin{eqnarray}
\left( \xi \overline \M \xi + \xi^\dagger \overline \M^\dagger \xi^\dagger
\right) &&= \left( \xi \M(\theta) \xi
+ \xi^\dagger \M^\dagger(\theta) \xi^\dagger \right) \nonumber\\
&&\quad+i {{a \overline \theta} \over {2b}} {1 \over \N}
\left( \Sigma - \Sigma^\dagger \right),
\end{eqnarray}
where $\M(\theta)$ and $\overline \theta$ are defined in Sect.~II.
The term proportional to $\overline \theta$ violates strong $CP$.

Planar QCD flavor symmetry constrains the coefficients of $\L^\M_{\rm baryon}$.
At leading order in the $1/\N$ expansion,
the coefficients of the singlet
perturbation are related
to the $a=3,8$ octet coefficients,
\begin{eqnarray}\label{massnonet}
\overline b^{0,1}_{(0)} &&= {1 \over \sqrt{6}} b^{0,8}_{(1)}, \nonumber\\
\overline b^{0,1}_{(2)} &&= {1 \over \sqrt{6}} \left( 2 b^{0,8}_{(2)}
+2 b^{0,8}_{(3)} \right).
\end{eqnarray}
The normalization of the singlet perturbation deviates from nonet symmetry
at relative order $1/\N$.

The final version of the leading $1/\N$ chiral Lagrangian containing explicit
symmetry breaking is as follows.
The Lagrangian
\begin{eqnarray}\label{lmfinal}
\L^{\M}_{\rm baryon} &&=
\Tr \left( \left( \xi \M(\theta) \xi
+ \xi^\dagger \M^\dagger(\theta) \xi^\dagger \right)
{\lambda^a \over 2} \right)
\H^a \nonumber\\
&& +{1 \over \N} \Tr \left( \left( \M(\theta) \Sigma
+ \M^\dagger(\theta) \Sigma^\dagger \right)
{I \over \sqrt{6}} \right) \H^0
\end{eqnarray}
for $a=3,8,9$, respects $CP$.
The strong-$CP$ violating Lagrangian is given by
\begin{eqnarray}\label{ltfinal}
\L^{\overline \theta}_{\rm baryon}
&&=i {{a \overline \theta} \over {2b}} {1 \over \N}
\Tr \left( \left( \Sigma - \Sigma^\dagger  \right)
{\lambda^a \over 2} \right)
\H^a \nonumber \\
&&i {{a \overline \theta} \over {2b}} {1 \over \N^2}
\Tr \left( \left( \Sigma - \Sigma^\dagger  \right)
{I \over \sqrt{6}} \right)
\H^0 \nonumber \\
\end{eqnarray}
for $a=3,8,9$.  Both of the these Lagrangians exhibit nonet
symmetry at leading order in the $1/\N$ expansion.  The second
terms in Eqs.~(\ref{lmfinal}) and~(\ref{ltfinal}) represent
planar QCD flavor breaking of relative
order $1/\N$ for the singlet perturbation.

\subsubsection{Comparison with Octet and Decuplet Chiral Lagrangian}

The quark mass terms of the $1/\N$ baryon chiral Lagrangian can
be compared to the quark mass terms of the octet and decuplet chiral
Lagrangian with no $1/\N$ expansion.  Strong $CP$ violation is neglected
in the following comparison.

To first order in the quark mass matrix, the chiral Lagrangian for
the octet and decuplet baryons is given by
\begin{eqnarray}
\L^{\M} &&=
\sigma\, \Tr \left( \M \left( \Sigma + \Sigma^\dagger \right) \right)
\Tr \left(\overline B B \right) \nonumber\\
&&-\tilde\sigma\, \Tr \left( \M \left( \Sigma + \Sigma^\dagger \right) \right)
\overline T^\mu T_\mu \nonumber\\
&& +b_D\, \Tr \overline B \left\{ \left( \xi^\dagger \M \xi^\dagger
+ \xi \M \xi \right), B \right\} \nonumber\\
&& +b_F\, \Tr \overline B \left[ \left( \xi^\dagger \M \xi^\dagger
+ \xi \M \xi \right), B \right] \nonumber\\
&& +c\,
\overline T^\mu \left( \xi^\dagger \M \xi^\dagger + \xi \M \xi \right) T_\mu ,
\end{eqnarray}
where $\sigma$ and $\tilde \sigma$ are the singlet quark mass parameters
of the octet and decuplet, respectively.  The parameters
$b_D$ and $b_F$ describe the flavor octet quark mass splittings of the baryon
octet, whereas the parameter $c$ describes the flavor octet quark
mass splittings of the baryon decuplet.

There is a one-to-one correspondence between the parameters of the octet
and decuplet chiral Lagrangian and the coefficients of the $1/\N$ baryon
chiral Lagrangian at $\N=3$.  The singlet quark mass parameters are related
to the $1/\N$ singlet coefficients by
\begin{eqnarray}\label{sst}
\sigma &&= {1 \over \sqrt{6}} \left( 3 b^{0,1}_{(0)}
+ {1 \over 4} b^{0,1}_{(2)} \right), \nonumber\\
\tilde \sigma &&= {1 \over \sqrt{6}} \left( 3 b^{0,1}_{(0)}
+ {5 \over 4} b^{0,1}_{(2)}\right) .
\end{eqnarray}
The octet quark mass parameters are related to the $1/\N$ octet coefficients
by
\begin{eqnarray}\label{bdbfc}
&&b_D = {1 \over 4} b^{0,8}_{(2)}, \nonumber\\
&&b_F = {1 \over 2}  b^{0,8}_{(1)} + {1 \over 6} b^{0,8}_{(2)}
+{1 \over {12}} b^{0,8}_{(3)},\\
&&c = - {3 \over 2} b^{0,8}_{(1)} - {5 \over 4} b^{0,8}_{(2)}
- {5 \over 4} b^{0,8}_{(3)}. \nonumber
\end{eqnarray}
Notice that the leading octet coefficient $b^{0,8}_{(1)}$ is pure $F$
and does not contribute to $b_D$.

Relations~(\ref{sst}) and~(\ref{bdbfc}) are valid for $\N$ set equal to
three.  For arbitrary $\N$, the $1/\N$ expansions of the singlet and octet
perturbations can be truncated for baryons with spins of order unity.
The leading singlet truncation
\begin{equation}
\H^0 = b^{0,1}_{(0)} \N \openone
\end{equation}
implies the parameter relation
\begin{equation}
\sigma = \tilde \sigma
\end{equation}
up to a correction of relative order $1/\N^2$.
For the $a=8$ perturbation, the leading truncation is
\begin{equation}\label{heighttrunc}
\H^8 = b^{0,8}_{(1)} T^8,
\end{equation}
up to a correction of order $1/\N$ since the mass splittings produced by
the operators $T^8$ and $\{J^i, G^{i8}\}$ are both order unity
in the $1/\N$ expansion\cite{jl}.  Eq.~(\ref{heighttrunc})
leads to the parameter relations
\begin{equation}\label{bcrel}
b_F = -{1 \over 3} c, \quad b_D=0,
\end{equation}
which are valid at order unity in the $1/\N$ expansion.
The subleading truncation
\begin{equation}
\H^8 = b^{0,8}_{(1)} T^8 + b^{0,8}_{(2)} {1 \over \N} \{ J^i, G^{i8} \}
\end{equation}
is valid up to a correction of order $1/\N^2$.  One linear combination
of the two parameter relations Eq.~(\ref{bcrel}) survives at this order,
\begin{equation}\label{relbc}
\left( b_D + b_F \right) = -{1 \over 3} c.
\end{equation}
The correction to this relation is order $1/\N^2$.
Eq.~(\ref{relbc}) is
in excellent agreement with the experimental values
$b_D m_s \sim 30\ {\rm MeV}$,
$b_F m_s \sim -95\ {\rm MeV}$,
and $2c m_s/3 \sim 150\ {\rm MeV}$
extracted in Ref.~\cite{jmasses}\footnote{A similar analysis applies for
the $\Delta S=1$ weak Lagrangian which is responsible for hyperon nonleptonic
decay.  The octet and decuplet $\Delta S=1$ weak Lagrangian involves three
parameters $h_D$, $h_F$ and $h_C$ which are in one-to-one correspondence
with the three coefficients of the $1/\N$ expansion for
$\H_{\rm weak}^{\Delta S =1}$\cite{djmtwo}.  The analogues of
Eqs.~(\ref{bcrel})
and~(\ref{relbc}) for $h_D$, $h_F$ and $h_C$ are obtained.
Eq.~(\ref{bcrel}) for hyperon nonleptonic decay was originally predicted
in the chiral quark model\cite{cqm}.
The experimental
values of these parameters extracted from the $s$-wave decays
at one-loop order in chiral perturbation
theory\cite{nonlep,springer} are consistent with these parameter relations.}.

Planar QCD flavor symmetry relates the singlet quark mass parameters
to the octet mass parameters,
\begin{eqnarray}\label{sigmanonet}
\sigma &&\rightarrow {1 \over 3}\left( 3 b_F - b_D \right), \nonumber\\
\tilde\sigma &&\rightarrow -{1 \over 3} c.
\end{eqnarray}
The consistency of Eq.~(\ref{sigmanonet}) with Eq.~(\ref{massnonet})
can be checked using Eqs.~(\ref{sst}) and~(\ref{bdbfc}).

\subsubsection{The Proton Matrix Element
$\left< p \left| m_s \overline s s \right| p \right>$}

Nonet symmetry amongst the linear quark mass splittings of the
baryons has implications for the analysis of the proton
matrix element $\left< p \left| m_s \overline s s \right| p \right>$.
The analysis of the linear in $m_s$ contribution to this matrix
element is discussed in this subsection.
The affect of contributions to the proton mass which are nonlinear
in quark masses can be computed using the methods of Sect.~IV.

The proton matrix element of the strange quark mass operator is
obtained by differentiation of the proton mass with respect to
$m_s$,
\begin{equation}
\left< p \left| m_s \overline s s \right| p \right> = m_s
{{\partial m_p} \over {\partial m_s}} .
\end{equation}
The standard chiral Lagrangian expansion of the proton mass to
linear order in the quark masses is
\begin{eqnarray}
m_p &&= m_B -2 \sigma \left( m_u + m_d + m_s \right)
+ 2 \left( b_F - b_D \right) m_s \nonumber\\
&&- 2 \left( b_F + b_D \right) m_u + {\rm nonlinear}\ldots,
\end{eqnarray}
which implies that
\begin{equation}\label{msss}
\left< p \left| m_s \overline s s \right| p \right> = 2 \left(
-\sigma + b_F - b_D \right) m_s
+ {\rm nonlinear} \ldots
\end{equation}
Substitution of the nonet symmetry relation Eq.~(\ref{sigmanonet})
leads to an exact cancellation of the $b_F$ term in the nonet symmetry
limit.

It is instructive to study this cancellation in the $1/\N$ expansion.
Expanding the proton mass to linear order in quark masses using the
$1/\N$ chiral Lagrangian, and differentiation with respect to $m_s$
leads to
\begin{eqnarray}
\left< p \left| m_s \overline s s \right| p \right> &&= -{2 \over\sqrt{6}}
\left( \N b^{0,1}_{(0)} + {3 \over 4} {1 \over \N} b^{0,1}_{(2)} \right)
m_s \nonumber\\
&&+{1 \over 3} \left( \N b^{0,8}_{(1)} + {3 \over 2}{1 \over \N} b^{0,8}_{(2)}
+ {3 \over 2}{1 \over \N} b^{0,8}_{(3)} \right) m_s \\
&&+ {\rm nonlinear} \ldots,\nonumber
\end{eqnarray}
where the relations
\begin{eqnarray}\label{t8g8}
T^8 &&= {1 \over {2\sqrt{3}}} \left( \N - 3 N_s \right), \nonumber\\
G^{i8} &&= {1 \over {2\sqrt{3}}} \left( J^i - 3 J^i_s \right),
\end{eqnarray}
have been used to evaluate proton matrix elements.  The nonet symmetry
conditions Eq.~(\ref{massnonet}) result in an exact cancellation amongst
the singlet and octet quark mass contributions.  Thus, the linear contribution
to $\left< p \left| m_s \overline s s \right| p \right>$ is produced entirely
by violation of nonet symmetry at order $1/\N$ in the $1/\N$ expansion,
\begin{eqnarray}\label{sbars}
\left< p \left| m_s \overline s s \right| p \right> &&=
O\left( {1 \over \N} \right) + {\rm nonlinear} \ldots,
\end{eqnarray}
where the $O(1/\N)$ term represents
$1/\N$-breaking of nonet symmetry in the singlet channel.
The above remarks generalize to arbitrary $\N$ if the proton
is identified with the strangeness-zero baryon of the spin-$1/2$
large-$\N$ flavor representation.  The $1/\N$ suppression
of Eq.~(\ref{sbars}) occurs because the proton contains no strange
quarks, so that the leading contribution to
$\left< p \left| m_s \overline s s \right| p \right>$ comes from
diagrams with a single quark loop (Fig.~5)\cite{luty}
in violation of planar QCD flavor symmetry.

It is conventional to rewrite Eq.~(\ref{msss}) in terms of the
sigma term
\begin{eqnarray}
\sigma_{\pi N} &&\equiv \hat m \left< p \left|
\overline u u +\overline d d \right| p \right> \nonumber\\
&&= -2 \hat m \left( 2 \sigma + b_D + b_F \right)
+ {\rm nonlinear}\ldots,
\end{eqnarray}
so that
\begin{eqnarray}
\left< p \left| m_s \overline s s \right| p \right> &&=
m_s \left( 3 b_F - b_D \right) + {1 \over 2} \left({m_s \over {\hat m}}
\right) \sigma_{\pi N} \nonumber\\
&&\quad+ {\rm nonlinear}\ldots
\end{eqnarray}
Nonet symmetry amongst the linear quark couplings implies a significant
cancellation between the first term and the sigma term.  This cancellation
explains sensitivity of central value of the proton matrix element
to the precise value of $\sigma_{\pi N}$.

\section{Nonanalytic Corrections}

The procedure for calculating non-analytic pion-loop corrections
using the $1/\N$ baryon chiral Lagrangian is examined in this section.
Aspects of this problem have been treated previously in
Refs.~\cite{dm,j,djm,ddjm,lm,lmw}.
The calculation of nonanalytic corrections to baryon amplitudes
in the $1/\N$ expansion at finite $\N$ introduces a number of issues
which have not been addressed before.
A sample calculation of the flavor $\bf 27$ nonanalytic
contribution to the baryon masses is presented in detail to illustrate
the technique.

The Feynman rules for baryon-pion couplings can be obtained from the
$1/\N$ chiral Lagrangian $\L_{\rm baryon}$ and $\L_{\rm baryon}^\M$.
The baryon propagator is given by inversion of the quadratic terms
in the Lagrangian.  This inversion is complicated by the presence of
the hyperfine and quark mass splittings.
In the chiral limit, $m_i \rightarrow 0$, the baryon propagator is
diagonal in spin, and can be written as
\begin{equation}\label{prop}
{ {i \P_\jmath} \over {\left( k^0 - \Delta_\jmath \right) } },
\end{equation}
where $\P_\jmath$ is a spin projection operator for spin $J=\jmath$
and
\begin{equation}
\Delta_\jmath = M_{\rm hyperfine}\Big\vert_{J^2 = \jmath (\jmath+1)}
-M_{\rm hyperfine}\Big\vert_{J^2 = \jmath_{\rm ext} (\jmath_{\rm ext}+1)}
\end{equation}
is the difference of the hyperfine mass splitting for spin $J=\jmath$
and the external baryon.
For $p$-wave pion emission, $\Delta_\jmath$ is given by
\begin{equation}
\Delta_\jmath = \cases{ {1 \over \N} 2 \jmath \ m^{0,1}_{(2)},
&$\jmath_{\rm ext} = \jmath -1$ \cr
\ 0, &$\jmath_{\rm ext} = \jmath$ \cr
-{1 \over \N}  2 \jmath \ m^{0,1}_{(2)},
&$\jmath_{\rm ext} = \jmath +1$ \cr }
\end{equation}
at leading order $1/\N$ in the $1/\N$ expansion, with subleading
terms beginning at order $1/\N^3$.
Eq.~(\ref{prop}) solves the inversion
problem in the chiral limit in terms of spin projection operators.

For arbitrary finite $\N$, the baryon tower consists of spins
$J=1/2,3/2,\ldots,\N/2$.  Each spin projection operator
must satisfy
\begin{eqnarray}
&&\P_\jmath^2 = \P_\jmath , \nonumber\\
&&\P_{\jmath^\prime} \P_\jmath =0, \qquad \jmath^\prime \ne \jmath,
\end{eqnarray}
by definition.
An explicit realization of these conditions is given by
\begin{equation}\label{proj}
\P_\jmath ={
{ \prod_{\jmath^\prime \ne \jmath}
\left( J^2 - J_{\jmath^\prime}^2 \right)} \over
{ \prod_{\jmath^\prime \ne \jmath}
\left( J_\jmath^2 - J_{\jmath^\prime}^2 \right) }
},
\end{equation}
where the
projection operator for spin $J_\jmath$ is given by
the product over all
$J_{\jmath^\prime} = 1/2,3/2,\ldots,\N/2$ not equal to $J_\jmath$.
For example, the spin-$1/2$ and $3/2$ projectors are given by
\begin{eqnarray}
\P_{1 \over 2} &&= {
{ \left( J^2 - { {15} \over 4} \right) \left( J^2 - { {35} \over 4} \right)
\ldots \left( J^2 - {\N \over 2} \left( {\N \over 2} +1 \right) \right) }
\over
{ \left( {3 \over 4} - { {15} \over 4} \right)
\left( {3 \over 4} - { {35} \over 4} \right)\ldots
\left( {3 \over 4} - {\N \over 2} \left( {\N \over 2} +1 \right)  \right) }
},
\nonumber\\
\P_{3 \over 2} &&= {
{ \left( J^2 - { {3} \over 4} \right) \left( J^2 - { {35} \over 4} \right)
\ldots \left( J^2 - {\N \over 2} \left( {\N \over 2} +1 \right) \right) }
\over
{ \left( {{15} \over 4} - { {3} \over 4} \right)
\left( {{15} \over 4} - { {35} \over 4} \right)\ldots
\left( {{15} \over 4} - {\N \over 2} \left( {\N \over 2} +1 \right)  \right) }
}.
\end{eqnarray}
Each of the projection operators Eq.~(\ref{proj}) is a
polynomial of degree $(\N-1)/2$ in $J^2$.

For $\N=3$, there are only two spins in the baryon tower.  The
spin projectors reduce to
\begin{eqnarray}
\P_{1\over 2} &&= - {1 \over 3} \left( J^2 - {{15} \over 4} \right),
\nonumber\\
\P_{3\over 2} &&= {1 \over 3} \left( J^2 - {3 \over 4} \right),
\end{eqnarray}
and the baryon propagator has the form Eq.~(\ref{prop}) with
\begin{eqnarray}
\Delta_{1\over 2} &&= \cases{0, &$\jmath_{\rm ext} = {1 / 2}$\cr
- \Delta, &$\jmath_{\rm ext} = {3 / 2}$\cr}\nonumber\\
\Delta_{3\over 2} &&= \cases{\Delta, &$\jmath_{\rm ext} = {1 / 2}$\cr
0, &$\jmath_{\rm ext} = {3 / 2}$\cr}
\end{eqnarray}
where
\begin{equation}
\Delta = {3 \over \N} m^{0,1}_{(2)}.
\end{equation}

Away from the chiral limit, quark mass splittings must be considered
in the inversion of the baryon quadratic terms.
Baryon mass splittings which are comparable to the pion octet masses
are to be retained in the baryon propagator.
For large-$\N$ baryons, the leading
hyperfine baryon mass splitting is order $\Lambda/\N$ whereas the
leading quark mass splittings are order $m_i$\footnote{The leading
$O(\N)$ terms in the quark mass perturbations $\H^a$, $a=0,3,8$
are proportional to the baryon identity operator $\openone$
and do not result in baryon mass splittings.  All $O(\N)$ mass
terms must be removed from the Lagrangian by the heavy field redefinition.}.
For QCD with $\N=3$,
these splittings satisfy the hierarchy
\begin{equation}
m_u,\ m_d \ll m_s < {\Lambda \over \N} \ .
\end{equation}
Only the leading quark mass splittings
proportional to $b^{0,8}_{(1)} T^8$ and the hyperfine mass splitting
are comparable to the pion octet masses in QCD.  Keeping these two
splittings amounts to the neglect
of isospin-breaking quark mass splittings and subleading quark mass splittings
of order $m_i/\N$.
The $T^8$
operator leads to spin-independent baryon mass splittings which are linear
in the number of strange quarks.
The baryon propagator is diagonal in spin and strange quark number
and is given by
\begin{equation}\label{proptwo}
{ {i \, \P_\jmath \, \P_{n_s}(\jmath) \, \P_\jmath} \over
{\left( k^0 - \Delta_\jmath - \Delta_{n_s} \right) } },
\end{equation}
where $\P_{n_s}(\jmath)$ is the $N_s = n_s$
strange quark projection operator of the spin-$\jmath$ flavor representation,
and (neglecting strong-$CP$ violation)
\begin{equation}
\Delta_{n_s} = {1 \over {2}} b^{0,8}_{(1)} (m_u + m_d - 2 m_s )
\left( n_s - n_{s \rm ext} \right),
\end{equation}
is the $T^8$ quark mass difference of the propagating
baryon and the external baryon.  Eq.~(\ref{proptwo})
solves the inversion problem in terms of strange quark projection operators.

The spin $J= \jmath$ large-$\N$
flavor representation contains baryons with
$N_s =0,1,\ldots,(\N+2 \jmath)/2$ strange quarks.
Strange-quark number projection operators
can be defined for the spin-$\jmath$ flavor representation in analogy
to the spin projection operators,
\begin{equation}
\P_{n_s}(\jmath) ={
{ \prod_{n_s^\prime \ne n_s}
\left( N_s - {n_s^\prime} \right)} \over
{ \prod_{n_s^\prime \ne n_s}
\left( {n_s} - {n_s^\prime} \right)}
},
\end{equation}
where the projection operator for $n_s$ strange quarks is given by the
product over all $n_s^\prime = 0,1,\ldots,(\N+2 \jmath)/2$ not equal to $n_s$.
For example, the zero and one strange quark projection operators
for the spin-$\jmath$ baryons are given by
\begin{eqnarray}
\P_{0}(\jmath) &&= {
{ \left( N_s - 1 \right) \left( N_s - 2 \right)
\ldots \left( N_s - \left({{\N+ 2 \jmath} \over 2}\right) \right) }
\over
{ \left( -1 \right)
\left( -2 \right)\ldots
\left( -\left( {{\N+ 2 \jmath} \over 2} \right) \right) }
},
\nonumber\\
\P_{1}(\jmath) &&= {
{ N_s \left( N_s - 2 \right)
\ldots \left(  N_s - \left({{\N+ 2 \jmath} \over 2}\right) \right) }
\over
{ \left( 1 \right)
\left( -1 \right)\ldots
\left( 1 -  \left({{\N+ 2 \jmath} \over 2}\right) \right) }
}.
\end{eqnarray}
Note that the projectors are different for each flavor representation
with a definite spin $J=\jmath$, since the allowed
strangeness sectors of a large-$\N$ flavor representation depends on
its spin.
Each strange-quark number
projection operator for spin $J=\jmath$ is a polynomial of
degree $(\N+2 \jmath)/2$ in $N_s$.

For $\N=3$, the spin-$1/2$ flavor representation contains baryons
with $0$, $1$, and $2$ strange quarks, while the spin-$3/2$ flavor
representation contains baryons with $0$, $1$, $2$, and $3$ strange
quarks.
The $J=1/2$ strange quark
projection operators reduce to
\begin{eqnarray}
\P_0({1 /2})
&&={1 \over 2}\left( N_s -1 \right) \left( N_s -2 \right), \nonumber\\
\P_1({1 / 2}) &&=-N_s\left( N_s -2 \right), \\
\P_2({1 / 2}) &&={1 \over 2}N_s \left( N_s -1 \right), \nonumber
\end{eqnarray}
whereas the $J=3/2$ strange quark projection operators reduce to
\begin{eqnarray}
\P_0({3 / 2}) &&=-{1 \over 6}\left( N_s -1 \right) \left( N_s -2 \right)
\left( N_s -3 \right), \nonumber\\
\P_1({3 / 2})
&&={1 \over 2}N_s\left( N_s -2 \right)\left( N_s -3 \right), \nonumber\\
\P_2({3 / 2}) &&=-{1 \over 2}N_s\left( N_s -1 \right)\left( N_s -3 \right), \\
\P_3({3 / 2})
&&={1 \over 6}N_s \left( N_s -1 \right) \left( N_s -2 \right). \nonumber
\end{eqnarray}
The baryon propagator has the form Eq.~(\ref{proptwo}) with $\Delta_{n_s}$
given by plus and minus
\begin{equation}
\Delta_{s} ={1 \over 2} b^{0,8}_{(1)} (m_u + m_d - 2 m_s )
\end{equation}
for $\Delta S= \pm 1$ transitions.

The generalization of the baryon propagator Eq.~(\ref{proptwo}) to include
all subleading quark mass splittings is provided in Appendix A for
completeness.

\subsection{Flavor-{\bf 27} Baryon Mass Splittings}

Flavor singlet and octet baryon mass splittings are present
in the $1/\N$ baryon chiral Lagrangian.  The flavor-$\bf 27$
mass splittings of the octet and decuplet
are calculable and nonanalytic in the quark masses and baryon
hyperfine mass splitting at leading order in chiral perturbation theory.
This mass splitting arises from the Feynman diagram Fig.~6.
The computation of the flavor-$\bf 27$
component of Fig.~6 for finite $\N$, $\N=3$, is presented
in detail in this section.
Computations at larger $\N$ are less interesting physically and more
complicated to extrapolate to $\N=3$ because
unphysical baryons participate as intermediate states in loop
diagrams\cite{dm,j},
and there are higher partial wave meson-baryon couplings which occur
at subleading orders\cite{djmtwo}.

The loop diagram Fig.~6 involves $\pi$, $K$ and $\eta$ emission and
reabsorption.  The $\eta^\prime$ meson is not included in the loop
since it is not soft relative to the baryons in QCD.
For degenerate heavy hadrons interacting with mesons,
the diagram Fig.~6 depends on a function $F(m)$ of the meson mass
$m$, which is obtained by performing the Feynman loop integral.
Neglecting isospin breaking, i.e. the $(m_d - m_u)$ quark mass difference,
the diagram depends on the function $F(m)$
for three meson mass values, $F(\pi)$,
$F(K)$, and $F(\eta)$, where the meson mass is denoted by its particle
label.  Any meson loop integral with
the exchange of a single meson,
in which a meson of flavor $a$ is emitted and a meson of flavor $b$
is reabsorbed,
can be written as a symmetric tensor with two adjoint (octet) indices
$a$ and $b$.  This symmetric tensor
decomposes into flavor singlet, adjoint ($\bf 8$) and
$\overline s s$ ($\bf 27$) representations,
\begin{eqnarray}\label{pitensor}
\Pi^{ab}&&=\frac 1 8 \left( 3 F(\pi) + 4 F(K) + F(\eta) \right)
\delta^{ab}\nonumber\\
&&+ \frac {2\sqrt{3}} 5
\left( \frac 3 2 F(\pi) - F(K) - \frac 1 2 F(\eta) \right) d^{ab8} \nonumber\\
&&+ \left( \frac 1 3 F(\pi) - \frac 4 3 F(K) + F(\eta) \right) \times \\
&&\qquad \left( \delta^{a8} \delta^{b8} - \frac 1 8 \delta^{ab}
- \frac 3 5 d^{ab8} d^{888} \right) \ , \nonumber
\end{eqnarray}
where the flavor singlet, octet and $\bf 27$ tensors in
Eq.~(\ref{pitensor}) are proportional to flavor singlet, octet and $\bf 27$
linear combinations of $F(\pi)$, $F(K)$ and $F(\eta)$.

For the Feynman diagram Fig.~6 with propagator $i/k^0$,
\begin{equation}
F(m) = {m^3 \over {24 \pi f_\pi^2} }.
\end{equation}
The flavor-$\bf 27$ combination of this function,
\begin{equation}
\left( \frac 1 3 F(\pi) - \frac 4 3 F(K) + F(\eta) \right),
\end{equation}
is highly suppressed relative
to the singlet and octet combinations, and is numerically very
small, of the order $4$~MeV.  For comparison, the flavor singlet combination
of this function is about $126$~MeV, while the flavor octet combination is
about $-38$~MeV.
This suppression of the flavor-$\bf 27$ combination is generic in chiral
perturbation theory.
It explains
the small violation of the Gell-Mann--Okubo formula for baryon
masses, which is $6.5$~MeV\footnote{This suppression mechanism was
originally noted in
the study of the octet and decuplet masses\cite{jmasses},
neglecting the pion mass.
The suppression of the flavor $\bf 27$ combination is
more significant when the pion mass is retained.}, as well as
the small flavor-$\bf 27$ mixing of the vector mesons\cite{vector}.
The suppression mechanism also applies to other meson-loop corrections
involving $\Pi^{ab}$, such as
flavor-$\bf 27$ chiral logarithmic
corrections to vertices with
\begin{equation}
F(m) = {m^2 \over {24 \pi^2 f_\pi^2}} \ln \left({ m^2 \over \mu^2 }\right).
\end{equation}
The flavor-$\bf 27$ chiral logarithm is numerically $0.035$.
Note that
the $\mu$-dependence of the chiral logarithm
cancels at leading order in the $\bf 27$ combination using the
Gell-Mann--Okubo formula for meson masses\cite{banerjee,martin}
\begin{equation}
\frac 1 3 m_\pi^2 - \frac 4 3 m_K^2 + m_\eta^2 =0.
\end{equation}

The computation of Fig.~6 is complicated significantly by the inclusion
of baryon mass splittings in the baryon propagator.  In this case, the
Feynman integral is a nonanalytic
function of the baryon mass splitting $\Delta$ as
well as the meson mass squared.
The function $F(m, \Delta)$ is defined
by the integral
\begin{eqnarray}\label{funcF}
i \,\delta^{ij}\  F(m, \Delta) &&={1 \over f^2}\int { {d^4k} \over
{(2\pi)^4} }\
{{i^2 ({\bf k}^i) (-{\bf k}^j)}\over
{(k^2 - m^2) (k^0 - \Delta) }}.
\end{eqnarray}
The precise formula for $F(m,\Delta)$ is given in Appendix B.
The computation of Fig.~6
is performed in this section using the baryon propagator Eq.~(\ref{prop}),
which neglects baryon flavor mass splittings, since the
generalization to the propagator Eq.~(\ref{proptwo})
can be obtained immediately from this formula.
With baryon propagator Eq.~(\ref{prop}),
the flavor-$\bf 27$ component of the meson tensor is given by
\begin{equation}
\Pi^{ab}_{27} =
\left( \delta^{a8} \delta^{b8} - \frac 1 8 \delta^{ab}
- \frac 3 5 d^{ab8} d^{888} \right)
I(\pi, K, \eta,\Delta),
\end{equation}
where
\begin{equation}\label{imd}
I(\pi,K,\eta,\Delta) = \left( \frac 1 3 F(\pi,\Delta)
- \frac 4 3 F(K, \Delta) + F(\eta, \Delta) \right) \ .
\end{equation}
Neglect of the $T^8$ quark mass splitting $\Delta_s$ affects
only the kaon loop graphs which involve $\Delta S= \pm1$
transitions.
The generalization of $I(\pi,K,\eta,\Delta)$ to
$I(\pi,K,\eta,\Delta,\Delta_s)$ is obtained by the replacement
\begin{equation}\label{idels}
F(K,\Delta) \rightarrow {1 \over 2} \left( F(K, \Delta+ \Delta_s)
+ F(K,\Delta-\Delta_s) \right)
\end{equation}
in Eq.~(\ref{imd}).

The diagram Fig.~6 is given by the product of a baryon operator
times the pion flavor tensor.
Using the baryon propagator Eq.~(\ref{prop}), Fig.~6 is given by
\begin{equation}\label{diagram}
{1 \over \N} \sum_\jmath \left( A^{ia} \P_\jmath A^{ib} \right)\
\Pi^{ab}_{27}\left( \Delta_\jmath \right),
\end{equation}
for arbitrary $\N$
where the baryon axial vector current operator $A^{ia}$ has a $1/\N$
expansion Eq.~(\ref{axin}).  The explicit factor of $1/\N$ occurs from
the rescaling $f_\pi \rightarrow \sqrt{\N} f$.
Eq.~(\ref{diagram}) reduces to
\begin{equation}\label{diagop}
{1 \over \N}\left( A^{i8} \P_{1 \over 2} A^{i8}
\ \, I(\Delta_{1\over 2})
+ A^{i8} \P_{3 \over 2} A^{i8}
\ \, I(\Delta_{3\over 2}) \right)
\end{equation}
for $\N=3$, where $A^{i8}$ has the $1/\N$ expansion Eq.~(\ref{aia}).
and the baryon operator is understood to be a $(0,{\bf 27})$, so that
subtraction of flavor singlet and octet components of
the baryon operator is implicit in the present notation.  The function
$I(\pi,K,\eta,\Delta_\jmath)$ is abbreviated as $I(\Delta_\jmath)$ in
Eq.~(\ref{diagop}).

The evaluation of Eq.~(\ref{diagop}) raises an important issue.
The baryon operator product of the two axial currents generates
$n$-body operators with $n> \N$ which are not operators in the
operator basis at finite $\N$.  In order to made sense of this
operator product, all of these higher body operators must be
rewritten as linear combinations of operators
in the operator basis with $n\le \N$.
Since the
operator basis is complete and independent\cite{djmtwo},
this reduction is always possible.
In practice, however, this operator reduction is formidable even
for the product of two axial vector currents.
The problem is solved in this work using spin projection operators.
The introduction of spin projection operators makes operator reduction
tractable and straightforward.
The details are presented in Appendix B.

There are two flavor-$\bf 27$ combinations of baryon masses,
the Gell-Mann--Okubo combination of octet baryon masses
\begin{equation}
{3 \over 4} \Lambda + {1 \over 4} \Sigma - {1 \over 2}
\left( N + \Xi \right),
\end{equation}
and the decuplet equal spacing rule combination\cite{jl},
\begin{equation}
-{4 \over 7} \Delta + {5 \over 7} \Sigma^* + {2 \over 7} \Xi^*
-{3 \over 7} \Omega \ .
\end{equation}
Violation of the Gell-Mann--Okubo formula is given by
\begin{equation}
{1 \over \N}\left( \P_{1 \over 2} A^{i8} \P_{1 \over 2} A^{i8} \P_{1 \over 2}
\ \,I(0)
+\P_{1 \over 2} A^{i8} \P_{3 \over 2} A^{i8} \P_{1 \over 2}
\ \,I(\Delta) \right),
\end{equation}
whereas violation of the flavor-$\bf 27$ equal spacing rule is given by
\begin{equation}
{1 \over \N}\left( \P_{3 \over 2} A^{i8} \P_{3 \over 2} A^{i8} \P_{3 \over 2}
\ \,I(0)
+\P_{3 \over 2} A^{i8} \P_{1 \over 2} A^{i8} \P_{3 \over 2}
\ \,I(-\Delta) \right) \ .
\end{equation}
Evaluation of the baryon operators yields
\begin{eqnarray}\label{gmo}
&&{3 \over 4} \Lambda + {1 \over 4} \Sigma - {1 \over 2}
\left( N + \Xi \right)
= \nonumber\\
&&\ {1 \over \N} \left[ \left(
{ 1 \over {16}} a_{1}^2
+ {3 \over 4} {1 \over \N} a_{1} b_{2}
+{9 \over {16}}{1 \over \N^2} b_{2}^2
+ {3 \over 8} {1 \over \N^2} a_{1} b_{3} \right. \right. \nonumber\\
&&\qquad\qquad +\left. {9 \over 4} {1 \over \N^3} b_{2} b_{3}
+ {9 \over {16}} {1 \over \N^4} b_{3}^2
\right) I(0)
\nonumber\\
&&\qquad +\left. \left(
{1 \over {8}} a_{1}^2 +{9 \over 8} {1 \over \N^2} a_{1} c_{3}
+ {{81} \over {32}} {1 \over \N^4} c_{3}^2
\right) I(\Delta)
\right],
\end{eqnarray}
and
\begin{eqnarray}\label{esr}
&&-{4 \over 7} \Delta + {5 \over 7} \Sigma^* + {2 \over 7} \Xi^*
-{3 \over 7} \Omega =\nonumber\\
&&\  {1 \over \N} \left[ \left(
{ {5} \over {8}} a_{1}^2
+ {{15} \over 4} {1 \over \N} a_{1} b_{2}
+{{45} \over 8}{1 \over \N^2} b_{2}^2
+ {{75} \over 4} {1 \over \N^2} a_{1} b_{3} \right.\right.\nonumber\\
&&\qquad\qquad + \left. {{225} \over 4} {1 \over \N^3} b_{2} b_{3}
+ {{1125} \over 8} {1 \over \N^4} b_{3}^2
\right) I(0)
\nonumber\\
&&\qquad -\left. \left(
{1 \over {4}} a_{1}^2 +{9 \over 4} {1 \over \N^2} a_{1} c_{3}
+{{81} \over {16}} {1 \over \N^4} c_{3}^2
\right) I(-\Delta)
\right],
\end{eqnarray}
where $a_1$, $b_2$, $b_3$ and $c_3$ are the coefficients of
the $1/\N$ expansion for the baryon axial vector current Eq.~(\ref{aia}).
Eqs.~(\ref{gmo}) and~(\ref{esr}) can be compared with the expressions
obtained in chiral perturbation theory with no $1/\N$ expansion\cite{jmasses},
\begin{equation}
\left(-{3 \over 4} \left( D^2 - 3 F^2 \right)
\ \overline I(0)
+ {1 \over 8} {\cal C}^2 \ \overline I(\Delta) \right)
\end{equation}
and
\begin{equation}
\left({{5} \over {18}} {\cal H}^2 \
\overline I(0)
- {1 \over 4} {\cal C}^2 \ \overline I(-\Delta) \right),
\end{equation}
respectively,
where the function $\overline I(\Delta) = I(\Delta) /\N$ is
proportional to $1/f_\pi^2$ rather than $1/f^2$.
Eqs.~(\ref{gmo}) and~(\ref{esr}) agree with these expressions
for $\N=3$, using the identifications Eq.~(\ref{dfch}).

Ref.~\cite{jl} showed that the two flavor-$\bf 27$ baryon mass splittings
are described by the $1/\N$ operators
\begin{equation}
c^{27,0}_{(2)} {1 \over \N} \{T^8, T^8 \}
+ c^{27,0}_{(3)} {1 \over \N^2} \{T^8, \{ J^i, G^{i8} \} \} \ ,
\end{equation}
so that one of the flavor-$\bf 27$ mass splittings is order $1/\N$
in the $1/\N$ expansion, whereas the other is order $1/\N^2$.  This
behavior is most easily seen for $\Delta=0$, where the $c^{27,0}_{(2)}$
mass combination is given by
\begin{eqnarray}
&&
5 \left({3 \over 4} \Lambda + {1 \over 4} \Sigma - {1 \over 2}
\left( N + \Xi \right) \right) - \left(
-{4 \over 7} \Delta + {5 \over 7} \Sigma^* + {2 \over 7} \Xi^*
-{3 \over 7} \Omega \right)\nonumber\\
&&= {9 \over {16}} {1 \over \N} a_1^2 \ I(\pi, K, \eta,0)
+ O\left({1 \over \N^3} \right),
\end{eqnarray}
while the $c^{27,0}_{(3)}$ mass combination is given by
\begin{eqnarray}\label{massii}
&&-2 \left({3 \over 4} \Lambda + {1 \over 4} \Sigma - {1 \over 2}
\left( N + \Xi \right)\right) + \left(
-{4 \over 7} \Delta + {5 \over 7} \Sigma^* + {2 \over 7} \Xi^*
-{3 \over 7} \Omega \right) \nonumber\\
&&={9 \over 4}{1 \over \N^2} a_1 b_2 \
I(\pi, K, \eta,0)
+ O\left({1 \over \N^3}\right),
\end{eqnarray}
so that
\begin{eqnarray}\label{dzerocoeff}
c^{27,0}_{(2)} &&= {1 \over 8} a_1^2 \ I(\pi,K,\eta,0)
+ O\left({1 \over \N^2} \right), \nonumber\\
c^{27,0}_{(3)} &&= {1 \over 2} a_1 b_2 \ I(\pi,K,\eta,0)
+ O\left({1 \over \N} \right).
\end{eqnarray}
Notice that there is no $a_1 b_2$ contribution to the $2$-body flavor-$\bf 27$
coefficient
and no $a_1^2$ contribution to $3$-body coefficient for $\Delta=0$.
For nonvanishing $\Delta$, it is still true that the first
flavor-$\bf 27$ mass splitting is order $1/\N$, while the second is
order $1/\N^2$.  The $1/\N$ counting is not explicit, however,
since $\Delta$ is implicitly order $1/\N$.
The expressions for the coefficients for $\Delta \ne 0$ are more
complicated,
\begin{eqnarray}
c^{27,0}_{(2)} &&= {1 \over 3} a_1^2 \
\left( -{5 \over {24}} I(0) + {5 \over {12}} I(\Delta) + {1 \over 6}
I(-\Delta) \right)
+ O\left({1 \over \N^2} \right), \nonumber\\
c^{27,0}_{(3)} &&=
{1 \over {18}} \N\, a_1^2
\ \left( 2 I(0) - I(\Delta) - I(-\Delta) \right)\\
&&\ +{1 \over 2} a_1 b_2 \ I(0) + O\left({1 \over \N} \right).\nonumber
\end{eqnarray}
As before, the coefficient $c^{27,0}_{(2)}$ does not
depend on $a_1 b_2$.  The coefficient $c^{27,0}_{(3)}$, however, now
appears to have an order $\N$ contribution proportional to $a_1^2$
which changes the $1/\N$ counting for this mass splitting.
This appearance is illusory.  Recall
that the function $I(\pi,K,\eta,\Delta)$ depends on the the meson
masses and $\Delta$ through the function $F(m, \Delta)$ defined in
Appendix B.  For $\vert \Delta \vert \le m$, the linear combination
$2 F(m,0) - F(m, \Delta) - F(m, -\Delta)$ is order $1/\N^2$, so the
term proportional to $a_1^2$ is $O(1/\N)$ and can be neglected
relative to leading $a_1 b_2$ term.  For $\vert \Delta \vert > m$,
the same linear combination reduces to $2 F(m,0)$, which is smaller
than an effect of order $1/\N^3$ since, by assumption, $m^3 < \Delta^3$.
Thus, the $3$-body coefficient reduces to the expression given in
Eq.~(\ref{dzerocoeff}) even for non-vanishing $\Delta$.

The theoretical calculation of the flavor-{\bf 27} mass
splittings can be compared with experiment.
The experimental value of the Gell-Mann--Okubo mass splitting is $6.53$~MeV
with negligible uncertainty.
The flavor-$\bf 27$ equal spacing rule mass
splitting depends on the unmeasured $\Delta^-$ mass which enters the isospin
zero mass
$\Delta_0 \equiv (\Delta^{++} + \Delta^{+} + \Delta^0 + \Delta^{-})/4$.
The $\Delta^-$ mass can be determined from the mass relation
\begin{equation}\label{delthree}
\Delta^{++} - 3 \Delta^+ + 3 \Delta^0 - \Delta^- =0,
\end{equation}
which is satisfied to order $1/\N^2$ in the $1/\N$ expansion and to
second order in isospin-breaking parameters\cite{jl}.  Numerically, the
$I=3$ mass difference of the $\Delta$ is at most of order $10^{-3}$~MeV,
so neglect of this mass difference introduces negligible error in
the determination of $\Delta^-$.
Using the value for $\Delta^-$ extracted with Eq.~(\ref{delthree})
and Particle Data Group values \cite{pdg} for the remaining $\Delta$
masses\footnote{There are three different measurements listed for
$\Delta^{++}$ and $\Delta^0$.  These measurements are averaged with
errors added in quadrature.}
yields $\Delta_0 = 1231.3 \pm 1.1$~MeV.  Evaluation of the flavor-$\bf 27$
equal spacing rule mass splitting using $\Delta_0$ and Particle Data Group
values for the remaining decuplet masses gives $6.75$~MeV, with an uncertainty
of $0.004$~MeV.

The theoretical
formulae depend on the two mass splittings $\Delta$ and $\Delta_s$, and
the flavor symmetric baryon-pion couplings $a_1$, $b_2$, $b_3$ and $c_3$
(or equivalently, $D$, $F$, ${\cal C}$ and ${\cal H}$).  The mass splittings
can be defined precisely\cite{jl}:
\begin{eqnarray}
\Delta &&= {1 \over {10}} \left( 4 \Delta_0 + 3 \Sigma_0^* + 2 \Xi^*_0 +
\Omega \right) \nonumber\\
&&\qquad - {1 \over 8} \left( 2 N_0 + 3 \Sigma_0 + \Lambda + 2 \Xi_0
\right)
\end{eqnarray}
and
\begin{eqnarray}
\Delta_s &&= {1 \over {10}} \left( \Delta_0 - {1 \over 2}\Xi_0^* -
{1 \over 2}\Omega \right) \nonumber\\
&&\qquad
- {1 \over 8} \left( 6 N_0 - 3 \Sigma_0 + \Lambda - 4 \Xi_0 \right),
\end{eqnarray}
where the zero subscripts refer to the $I=0$ mass combinations defined in
Ref.~\cite{jl}.  Evaluation of the mass splittings yields
$\Delta = 230.7 \pm 0.1$~MeV and $\Delta_s = 225.7 \pm 0.03$~MeV.  The
baryon axial couplings were extracted from experiment
in Ref.~\cite{ddjm}.
The Gell-Mann--Okubo mass combination depends primarily on the coefficients
$a_1$ and $b_2$, which are fairly well-determined.
The flavor-{\bf 27} equal spacing rule mass combination, however, is
sensitive to the value of $b_3$ which is difficult to extract
from experiment\footnote{Because of this uncertainty, one could consider the
alternative of extracting the baryon
axial couplings from the flavor-{\bf 27} mass splittings, as suggested
recently in Ref.~\cite{banerjee}.  However, the theoretical formula receives
sizeable (unknown) corrections at higher orders in chiral perturbation theory,
so error bars on the extracted couplings are significant.}.
This sensitivity is due to the large numerical constants
of the $b_3$ terms appearing in Eq.~(\ref{esr}), which in turn
is a reflection of the fact that the spin-$3/2$ flavor representation is
at the top of the baryon tower for $\N=3$ so
the presumption $J/\N \sim O(1/\N)$ is breaking down.

Besides the uncertainty of the baryon axial couplings, numerical
evaluation of the theoretical formulae for the
flavor-{\bf 27} mass splittings is further complicated by the sensitivity
of the numerics to the precise formulae which are used.  For
example, imposition of the Gell-Mann--Okubo formula for the meson masses
changes the numerical value of the function $I(\pi,K,\eta,\Delta)$
considerably.  In addition, the generalization of the theoretical formulae
to include $\Delta_s$ through Eq.~(\ref{idels}) changes the numerics
significantly.  Note that for non-vanishing
$\Delta_s$, the function $I(\Delta)$ depends on $\Delta_s$
through Eq.~(\ref{idels}) even for $\Delta=0$.
Although the $\mu$-dependence of the flavor-{\bf 27} combination of
$F(m, \Delta,\mu)$ cancels at leading order using the Gell-Mann--Okubo formula
for the meson masses if $\Delta_s=0$, there is additional $\mu$-dependence
from the term $\Delta \Delta_s^2 \ln m_K^2/ \mu^2$ when
$\Delta_s \ne 0$ and $\Delta \ne 0$.  This
$\mu$-dependence is cancelled by a finite counterterm proportional to
$\Delta \Delta_s^2$.  At leading order in chiral perturbation theory,
it is possible
to drop the $\Delta \Delta_s^2 \ln m_K^2/ \mu^2$ chiral logarithm
from the function $I$ to obtain a $\mu$-independent quantity.  It is
also possible to drop the finite $\Delta \Delta_s^2$ and $\Delta_s m^2$ terms
in $I$.
Note that
$\Delta_s$ still appears in the last term of Eq.~(B1).

The mass combination in Eq.~(\ref{massii}) does not depend sensitively on
$c_3$ or on the $\mu$-dependence of the $\Delta \Delta_s^2$
chiral logarithm.
Numerical evaluation
of Eq.~(\ref{massii}) using $a_1 = 2(0.791 \pm 0.007)$ and
$b_2 = 6(-0.058 \pm 0.011)$, as extracted in Ref.~\cite{ddjm}, yields
$-5.3$~MeV,
whereas the experimental value is $-6.2$~MeV.  The agreement is consistent with
a correction of relative order $1/\N$.

\section{Planar QCD Flavor Symmetry and $SU(3)$ Flavor Symmetry
Breaking}

The two-flavor version of planar QCD flavor symmetry has implications
for the structure of $SU(3)$ symmetry breaking at leading order in $1/\N$.
Planar QCD for three light flavors exhibits an approximate
$U(3)$ flavor symmetry, which is broken explicitly by the quark mass
matrix.  In the planar limit,
the flavor symmetry breaking due to the quark mass
matrix transforms as the $a=3,8,9$ components of a nonet.
Neglecting isospin breaking, there is an
unbroken $U(2)$ flavor symmetry, which includes the diagonal generators
$\sigma^3$ and $I$ in the $2 \times 2$ subspace.  This residual
$U(2)$ flavor symmetry can
constrain the form of $SU(3)$-breaking
in the $1/\N$ expansion.  The constraints must
be satisfied
at each and every order in $SU(3)$ flavor symmetry
breaking.

The relevance of $U(2)$ symmetry for $SU(3)$
breaking is illustrated by the baryon axial vector currents.
The $F=2$ version of planar QCD flavor symmetry constrains the
leading coefficients of the isosinglet axial vector current $A^i$
relative to the coefficients of the isovector axial current $A^{ia}$,
$a =1,2,3$,
\begin{eqnarray}\label{twoconstraint}
\overline b^{1,1}_{(1)} &=& {1 \over 2} \left( a^{1,3}_{(1)}
+ b^{1,3}_{(2)}\right), \nonumber\\
\overline b^{1,1}_{(3)} &=& {1 \over 2} \left( 2 b^{1,3}_{(3)}\right).
\end{eqnarray}
Eq.~(\ref{twoconstraint}) is the two-flavor analogue of
Eq.~(\ref{axiconstraint}).  The isosinglet and isovector axial
currents couple to the pion quartet
$\bbox{\Phi} = \pi^a {\sigma^a \over 2} + \tilde\eta {I \over {2}}$,
where $\tilde \eta$ is an admixture of $\eta$ and $\eta^\prime$.
The $\eta$ ($\eta^\prime$) couplings of baryons with zero strangeness
are each proportional to the $\tilde \eta$ couplings.
Thus, the $\eta$ ($\eta^\prime$) couplings of strangeness-zero
baryons are normalized relative to the pion couplings
in the presence of $SU(3)$ breaking by Eq.~(\ref{twoconstraint})
at leading order in $1/\N$.

Ref.~\cite{djmtwo} derived the flavor-octet
baryon axial vector current to linear
order in $SU(3)$ symmetry breaking $\epsilon$ and leading order
in $1/\N$.
The first constraint of Eq.~(\ref{twoconstraint}) can
be imposed on this current in the strangeness-zero sector.
For strangeness-zero baryons, the $1/\N$ expansion of the
baryon axial vector current
to linear order in $SU(3)$ symmetry breaking reduces to
\begin{eqnarray}\label{aplusda}
A^{ia} &&+ \delta A^{ia} = \left( a  G^{ia} + b {1 \over \N}
J^i T^a \right) \nonumber\\
&&+\epsilon d^{ab8} \left( c_1 G^{ib} + c_2 {1 \over \N}
J^i T^b \right) + \epsilon c_6 \delta^{a8} J^i,
\end{eqnarray}
up to terms of relative order $1/\N^2$.
Note that the
coefficients $a$ and $b$ contain
contributions of order $\epsilon$ and reduce to $a_1$ and $b_2$ only
in the absence of $SU(3)$ breaking.
The coefficients $a$ and $b$ automatically satisfy the first constraint
of Eq.~(\ref{twoconstraint}) due to the $SU(3)$ symmetry of their operators.
The remaining terms contribute the following to the coefficients of
the $F=2$ isosinglet and isovector $1/\N$ expansions
when evaluated for strangeness-zero baryons,
\begin{eqnarray}
\delta b_{(1)}^{1,1} &&= \sqrt{3} \epsilon \left( - {1 \over 6}
\left( c_1 + c_2\right) + c_6 \right), \nonumber\\
\delta a^{1,3}_{(1)} && ={1 \over \sqrt{3}} \epsilon c_1,\\
\delta b^{1,3}_{(2)} && ={1 \over \sqrt{3}} \epsilon c_2. \nonumber
\end{eqnarray}
The factor of $\sqrt{3}$ in the
first equation occurs because $\tilde \eta$ is proportional to
$\eta/ \sqrt{3}$.  The
first quartet symmetry constraint of Eq.~(\ref{twoconstraint}) implies that
\begin{equation}
\overline c_6 = {1 \over 3} \left( c_1 + c_2 \right),
\end{equation}
up to a correction of relative order $1/\N$.  This is the same
constraint on $c_6$ obtained in
Ref.~\cite{djmtwo}.  The above derivation shows that this constraint
follows from $F=2$ planar QCD flavor symmetry.

The second constraint in Eq.~(\ref{twoconstraint}) applies to
spin-diagonal order $1/\N^2$ terms which have not been included
in Eq.~(\ref{aplusda}).
These neglected terms reduce to
\begin{eqnarray}
&&b^\prime {1 \over \N^2} \{J^i, \{J^j, G^{ja} \}\}
\nonumber\\
&&+\epsilon d^{ab8} d_1
{1 \over \N^2} \{J^i, \{J^j, G^{jb} \}\}
+ \epsilon c_7 \delta^{a8} {1 \over \N^2}\{J^2, J^i\},
\end{eqnarray}
in the strangeness-zero sector, where
the coefficient $b^\prime$ is equal to $b_3$ in the $SU(3)$
flavor symmetry limit, but contains a contribution of order
$\epsilon$ at linear order in $SU(3)$ breaking.
The coefficient $b^\prime$ automatically
satisfies the second constraint of Eq.~(\ref{twoconstraint})
due to the $SU(3)$ symmetry of the operator.
The remaining terms are constrained to satisfy
\begin{equation}
\overline c_7 ={2 \over 3} d_1
\end{equation}
up to a correction of relative order $1/\N$.

The above analysis shows that
$SU(3)$
flavor symmetry breaking does not alter the relationship between
pion and $\eta$ couplings of strangeness-zero baryons from exact
$SU(3)$ flavor symmetry to leading order in $1/\N$.  Any
violation of this $SU(3)$ normalization requires $SU(3)$
flavor symmetry breaking in a $1/\N$-suppressed quark loop.
This result was originally reported in Ref.~\cite{djmtwo}.

\section{Conclusions}

A $1/\N$ chiral Lagrangian for baryons is formulated
which correctly implements planar QCD flavor symmetry and
$({\rm spin} \otimes {\rm flavor})$ symmetry for baryons.
The constraints
of planar QCD flavor symmetry on the baryon $1/\N$ expansion
have not been realized previously, and are presented
in detail in this work.
These constraints
are valid to leading order
in the $1/\N$ expansion operator by operator in the
baryon $1/\N$ expansion.  Thus, planar QCD flavor
symmetry constrains the operator coefficients at leading
order in $1/\N$.  The symmetry implies that baryon flavor octet
and singlet amplitudes form nonets at leading order in $1/\N$.
Specific examples of nonet symmetry include the formation
of a flavor nonet axial vector current from the flavor singlet
and flavor octet axial vector baryon currents at leading order in
$1/\N$, as well as the formation of a nonet amongst the flavor singlet
and octet baryon mass terms with linear dependence on the quark masses $m_i$.

The formulation of the baryon chiral Lagrangian in terms of operators
with definite $1/\N$ dependence enables one to study the precise
$1/\N$ structure of the chiral expansion.  In specific instances, such as
the proton matrix element
$\left< p \left| m_s \overline s s \right| p \right>$,
the leading $1/\N$ terms cancel exactly, so the $1/\N$ expansion explains
the suppression of the quantity.

The calculation of non-analytic meson-loop corrections in $1/\N$ baryon
chiral perturbation theory at finite $\N$ is addressed.  The
$1/\N$ and group theoretic structure of the loop corrections is
manifest using the method described in this work.
The introduction of spin projection operators
simplifies the formidable problem of operator reduction, making
calculations tractable.
A specific
example of the flavor-$\bf 27$ meson-loop contribution
to the baryon mass splittings is presented in
detail.
The $1/\N$ computation provided in this work
generalizes the formulae obtained previously
in ordinary baryon chiral perturbation theory with $\N=3$
to include the leading
flavor octet mass splitting $\Delta_s$ of the baryons.
The $1/\N$ formulae reveal the $1/\N$ and flavor-breaking
structure of the
flavor-$\bf 27$ baryon mass splittings at leading order in
chiral perturbation theory.

\section{Acknowledgments}
I am grateful to A. Manohar for useful discussions.
This work was supported by the Department of Energy
under grant DOE-FG03-90ER40546; by the NYI program,
through Grant No. PHY-9457911 from the National Science Foundation;
and by a research fellowship from the Alfred P. Sloan Foundation.

\vfill\break\eject

\appendix
\section{Baryon Propagator}
The generalization of the baryon propagator Eq.~(\ref{proptwo})
to include all subleading quark mass splittings is provided
in this appendix for
completeness.  Isospin-breaking quark mass splittings are neglected in the
discussion for simplicity.  These splittings can be included at the expense
of introducing additional projection operators.

Including all hyperfine and $a=0,8$ quark
mass splittings, the baryon propagator is given by
\begin{equation}\label{propthree}
{ {i \, \P_\jmath \, \P_{n_s}(\jmath) \, \P_\imath(\jmath,n_s)
\,\P_{n_s}(\jmath) \, \P_\jmath} \over
{\left( k^0 - \Delta_\jmath - \Delta_{0} - \Delta_{8} \right) } },
\end{equation}
where $\P_\imath(\jmath,n_s)$ is the projection operator for isospin
$I=\imath$ in the $N_s = n_s$ strange quark sector of the spin $J=\jmath$
flavor representation, and $\Delta_0$ and $\Delta_8$ are the $a=0$ and
$a=8$ quark mass splittings of the propagating baryon relative to the
external baryon.  The $a=0$ baryon quark mass operator
\begin{equation}
M^0 = -{2 \over \sqrt{6}} \left( m_u + m_d + m_s \right) \H^0
\end{equation}
depends only on polynomials of $J^2$, so
\begin{equation}
\Delta_0 = M^0\Big\vert_{J^2 = \jmath (\jmath+1)}
- M^0\Big\vert_{J^2 = \jmath_{\rm ext} (\jmath_{\rm ext}+1)}
\end{equation}
is already diagonalized by the spin projection operators.  The $a=8$
baryon quark mass operator
\begin{equation}
M^8 = -{1 \over \sqrt{3}} \left( m_u + m_d -2 m_s \right) \H^8
\end{equation}
involves two operator series, generated by $T^8$ and $\{J^i, G^{i8}\}$
times polynomials in $J^2$.  The operator series involving $T^8$ times
polynomials in $J^2$ is diagonalized by the spin and strange quark number
projection operators.  The operator series involving $\{J^i, G^{i8} \}$
requires the introduction of isospin projection operators.

For arbitrary $\N$, the structure of the baryon multiplets is such that:
$(i)$ the isospin of a baryon is equal to the total angular momentum
(spin) of the up and down
quarks, $I= J_{ud}$, and $(ii)$ the total angular momentum (spin) of the
strange quarks is equal to one-half the number of strange quarks,
$J_s = N_s/2$.  Since $J= J_{ud} + J_s$, it therefore follows that
spin-$\jmath$
baryons can only have isospins $I = \vert {N_s \over 2 } + \jmath \vert,
\vert {N_s \over 2 } + \jmath \vert -1, \ldots,
\vert {N_s \over 2 } - \jmath \vert$, where all possible isospins are
allowed for $2\jmath \le N_s \le (\N - 2\jmath)/2$, but only
a subset of the isospins are allowed for $0 \le N_s < 2 \jmath$ and
$(\N -2 \jmath )/2 < N_s \le (\N + 2 \jmath)/2$.
For $N_s=0$, only the largest isospin is allowed.  For $N_s =1$, only
the two largest isospins are allowed.  This pattern of one additional
allowed isospin as $N_s$ increases by one unit continues for
the interval $0 \le N_s < 2 \jmath$ until the full set of isospins is
allowed for $N_s = 2 \jmath$.  Similarly, for $N_s = (\N + 2 \jmath)/2$,
only the smallest isospin is allowed.  For $N_s = (\N + 2 \jmath)/2 -1$,
only the two smallest isospins are allowed.  This pattern of
one additional allowed isospin as $N_s$ decreases by one unit continues for
$(\N -2 \jmath )/2 < N_s \le (\N + 2 \jmath)/2$ until the full set of
isospins is allowed for $N_s =(\N -2 \jmath )/2$.

It is easier to digest this pattern of isospins if one specializes
to the spin-$1/2$ and spin-$3/2$ flavor representations with the weight
diagrams displayed in Figs.~3 and~4.
For $J=1/2$, there are two allowed isospins $I= (N_s +1)/2$ and
$I= (N_s -1)/2$ for $1 \le N_s \le (\N-1)/2$.  Both $N_s =0$ and
$N_s = (\N+1)/2$ are exceptions:  for $N_s =0$, $I=1/2$, whereas
for $N_s = (\N+1)/2$, $I = (\N-1)/4$.
For $J=3/2$, there are four
allowed isospins $I= (N_s +3)/2, (N_s +1)/2, (N_s -1)/2, (N_s -3)/2$
for $3 \le N_s \le (\N-3)/2$.  The three smallest
and largest strangeness sectors $N_s =0,1,2$ and
$N_s = (\N-1)/2$, $(\N+1)/2$ and $(\N+3)/2$ are special cases:
for $N_s =0$, there is one allowed $I=3/2$; for $N_s = 1$, there are
two allowed isospins, $I=2$ and $I=1$; for $N_s=2$, there are three
allowed isospins $I= 5/2,3/2,$ and $1/2$; while for $N_s = (\N-1)/2$,
there are three allowed isospins, $I= (\N -7)/4$, $(\N - 3)/4$, and
$(\N +1)/4$;
for $N_s = (\N+1)/2$, there
are two allowed isospins $I=(\N -5)/4$ and $(\N -1)/4$; and for
$N_s = (\N + 3)/2$, there is one allowed isospin
$I = (\N -3)/4$.

The remaining components of the baryon propagator Eq.~(\ref{propthree})
can now be defined.
Isospin projection
operators for the spin-$\jmath$ baryons with $n_s$
strange quarks are given by
\begin{equation}\label{iproj}
\P_\imath(\jmath, n_s)={
{ \prod_{\imath^\prime \ne \imath}
\left( I^2 - I_{\imath^\prime}^2 \right)} \over
{ \prod_{\imath^\prime \ne \imath}
\left( I_\imath^2 - I_{\imath^\prime}^2 \right) }
},
\end{equation}
where the projection operator for isospin $I_\imath$ is
given by the product over all $I_{\imath^\prime}$ not equal to $I_\imath$.
For $2 \jmath \le n_s \le (\N -2 \jmath)/2$,
$I_{\imath^\prime}= \vert (n_s + 2 \jmath)/2 \vert$,
$\vert (n_s + 2 \jmath)/2 \vert -1$, $\ldots$$\vert (n_s - 2 \jmath)/2 \vert$.
The allowed isospins for the smallest and largest $n_s$ values
of the spin-$\jmath$
flavor representations vary according the pattern described above.
The operator series involving $\{J^i, G^{i8} \}$ depends on the
operator $J \cdot J_s = {1 \over 2} \left( J^2 + J_s^2 - I^2 \right)$,
which in turn depends on isospin.  It is easy to evaluate $J \cdot J_s$
for each of the allowed isomultiplets of the spin-$\jmath$ flavor
representation.  For example, for the spin-$1/2$ representation,
$J\cdot J_s$ is equal to $-N_s/4$ for
isomultiplets with $I=(N_s+1)/2$, and $(N_s+2)/4$ for isomultiplets with
$I= (N_s-1)/2$.
For spin-$3/2$, $J \cdot J_s$ equals $-3N_s/4$, $-(N_s -6)/4$, $(N_s +8)/4$
and $3(N_s +2)/4$ for the
$I=(N_s+3)/2$, $(N_s+1)/2$, $(N_s-1)/2$ and
$(N_s -3)/2$ isomultiplets, respectively.  Thus, the baryon mass difference
\begin{equation}
\Delta_8 =
M^8 \Big\vert_{\jmath, n_s, \imath}-
M^8 \Big\vert_{\jmath_{\rm ext}, n_{s\, {\rm ext}}, \imath_{\rm ext}}
\end{equation}
is diagonalized by the spin, strange quark number and isospin projection
operators.

\onecolumn
\widetext
%
%
\section{Flavor-${\bf 27}$ Nonanalytic Mass Splittings}

This appendix provides additional formulae for the
calculation of the flavor-$\bf 27$ baryon mass splittings from
Fig.~6.  The function $F(m,\Delta)$ defined in Eq.~(\ref{funcF})
is given by \cite{jmhungary}
\begin{equation}
{24 \pi^2 f^2} \ F(m, \Delta, \mu) = \cases{
\Delta \left(\Delta^2 -{3 \over 2} m^2 \right)
\ln{m^2 \over \mu^2}- {8 \over 3} \Delta^3-{7 \over 2}\Delta m^2
+ 2 \left( m^2 - \Delta^2 \right)^{3/2} \left[ {\pi \over 2} - {\rm tan}^{-1}
\left( {\Delta \over \sqrt{m^2 - \Delta^2}}\right) \right],
& $\left|\Delta\right| \le m$,\cr
\Delta \left(\Delta^2 -{3 \over 2} m^2 \right)
\ln{m^2 \over \mu^2}- {8 \over 3} \Delta^3-{7 \over 2}\Delta m^2
- \left( \Delta^2 - m^2 \right)^{3/2}
\ln \left( { {\Delta - \sqrt{\Delta^2 - m^2}} \over
{\Delta + \sqrt{\Delta^2 - m^2}}} \right)
,& $\left|\Delta\right| > m$ ,\cr}
\end{equation}
where finite terms with mass dependence $\Delta^3$ and $m^2 \Delta$
have been retained.  For $\Delta_s \ne 0$, there is a chiral logarithmic
contribution to the flavor-{\bf 27} combination proportional to
$\Delta \Delta_s^2$, as well as a finite term.
For $\Delta=0$,
the function reduces to
\begin{equation}
F( m, 0) = {m^3 \over {24 \pi f^2}} \ .
\end{equation}

The remainder of the appendix is devoted to evaluating the flavor-$\bf 27$
baryon operator
\begin{equation}
{1 \over \N}\left( A^{i8} \P_{1 \over 2} A^{i8}
\ \, I(\Delta_{1\over 2})
+ A^{i8} \P_{3 \over 2} A^{i8}
\ \, I(\Delta_{3\over 2}) \right)\ ,
\end{equation}
using spin projection operators.

For $\N=3$, the baryon axial current $A^{i8}$ has
a $1/\N$ expansion in terms of the four operators of Eq.~(\ref{aia}).
The flavor-$\bf 27$ baryon operator product contains $n$-body
operators, $n > \N$, which are complicated anticommutators of
the $1$-body operators.  Each of these operators contains two flavor octet
$1$-body operators, which each may be either $T^8$ or $G^{i8}$; all remaining
$1$-body operators in the operator product are spin operators.  Operator
reduction of these spin operators is immediately possible using spin
projection operators.  The following observations facilitate this
reduction:
\begin{itemize}
\item $A^{i8} \P_\jmath A^{i8}$ is purely diagonal.
\item $J^i$ is purely diagonal.
\item $\{J^i, G^{i8} \}$ is purely diagonal, and
$\{J^i, G^{i8} \} = 2 J^i G^{i8} = 2 G^{i8} J^i$ since $[J^i, G^{i8}]=0$.
\item
$\left(\{J^2, G^{i8} \} - {1 \over 2} \{ J^i, \{ J^j, G^{j8} \}\} \right)$
is purely off-diagonal, since the second term
subtracts off the diagonal component of $\{J^2, G^{i8} \}$.  Thus, this
operator reduces to
$\P_{1 \over 2} \{ J^2, G^{i8} \} \P_{3 \over 2}
+\P_{3 \over 2} \{ J^2, G^{i8} \} \P_{1 \over 2}$, which
can be replaced by
${9 \over 2} \left( \P_{1 \over 2} G^{i8} \P_{3 \over 2}+
\P_{3 \over 2} G^{i8} \P_{1 \over 2}\right)$.
\end{itemize}

Using the above observations, it is straightforward to show that
\begin{eqnarray}\label{halfint}
{1 \over \N} A^{i8} \P_{1 \over 2} A^{i8} &&=
{1 \over \N}
\P_{1 \over 2} G^{i8} \P_{1 \over 2} G^{i8} \P_{1 \over 2}
\left( a_{1}^2 + 6 {1 \over \N^2} a_{1} b_{3} +
9 {1 \over \N^4} b_{3}^2 \right)
\nonumber\\
&&+ {1 \over \N} \P_{3 \over 2} G^{i8} \P_{1 \over 2} G^{i8} \P_{3 \over 2}
\left( a_{1}^2 + 9 {1 \over \N^2} a_{1} c_{3} +
\left({9 \over 2} \right)^2 {1 \over \N^4} c_{3}^2 \right)
\nonumber\\
&&+ {1 \over \N} \P_{1 \over 2} T^8 \P_{1 \over 2} T^8 \P_{1 \over 2}
\left({3 \over 4} {1 \over \N^2} b_{2}^2 \right)
\nonumber\\
&&+{1 \over \N^2} \left( \P_{1 \over 2} G^{i8} \P_{1 \over 2} J^i T^8
\P_{1 \over 2}
+ \P_{1 \over 2} J^i T^8 \P_{1 \over 2} G^{i8} \P_{1 \over 2} \right)
\left( a_{1} b_{2} + 3 {1 \over \N^2} b_{2} b_{3} \right),
\end{eqnarray}
and
\begin{eqnarray}\label{thalfint}
{1 \over \N} A^{i8} \P_{3 \over 2} A^{i8} &&=
{1 \over \N}
\P_{3 \over 2} G^{i8} \P_{3 \over 2} G^{i8} \P_{3 \over 2}
\left( a_{1}^2 + 30 {1 \over \N^2} a_{1} b_{3} +
225 {1 \over \N^4} b_{3}^2 \right)
\nonumber\\
&&+ {1 \over \N} \P_{1 \over 2} G^{i8} \P_{3 \over 2} G^{i8} \P_{1 \over 2}
\left( a_{1}^2 + 9 {1 \over \N^2} a_{1} c_{3} +
\left({9 \over 2} \right)^2 {1 \over \N^4} c_{3}^2 \right)
\nonumber\\
&&+ {1 \over \N} \P_{3 \over 2} T^8 \P_{3 \over 2} T^8 \P_{3 \over 2}
\left({{15} \over 4} {1 \over \N^2} b_{2}^2 \right)
\nonumber\\
&&+{1 \over \N^2} \left( P_{3 \over 2} G^{i8} \P_{3 \over 2} J^i T^8
\P_{3 \over 2}
+ \P_{3 \over 2} J^i T^8 \P_{3 \over 2} G^{i8} \P_{3 \over 2} \right)
\left( a_{1} b_{2} + 15 {1 \over \N^2} b_{2} b_{3} \right).
\end{eqnarray}
The flavor-$\bf 27$ mass combinations,
the Gell-Mann--Okubo combination of spin-$1/2$ octet baryons,
\begin{equation}
{3 \over 4} \Lambda + {1 \over 4} \Sigma - {1 \over 2}
\left( N + \Xi \right)
\end{equation}
and the equal spacing rule of the spin-$3/2$ baryons,
\begin{equation}
-{4 \over 7} \Delta + {5 \over 7} \Sigma^* + {2 \over 7} \Xi^*
-{3 \over 7} \Omega
\end{equation}
are given by
\begin{equation}\label{halfop}
-{1 \over \N}\left( \P_{1 \over 2} A^{i8} \P_{1 \over 2} A^{i8} \P_{1 \over 2}
\ \,I(0)
+\P_{1 \over 2} A^{i8} \P_{3 \over 2} A^{i8} \P_{1 \over 2}
\ \,I(\Delta) \right),
\end{equation}
and
\begin{equation}\label{thalfop}
-{1 \over \N}\left( \P_{3 \over 2} A^{i8} \P_{3 \over 2} A^{i8} \P_{3 \over 2}
\ \,I(0)
+\P_{3 \over 2} A^{i8} \P_{1 \over 2} A^{i8} \P_{3 \over 2}
\ \,I(-\Delta) \right),
\end{equation}
respectively, where the overall minus signs occur because the
baryon mass term in the Lagrangian is negative.
Each of the baryon operators appearing in
Eqs.~(\ref{halfop}) and~(\ref{thalfop}) can be obtained from
Eqs.~(\ref{halfint}) and~(\ref{thalfint}).

The $(0,{\bf 27})$ operator basis consists of two operators, $\{T^8,T^8\}$
and $\{T^8, \{J^i, G^{i8}\}\}$.  The above reduction has left one additional
operator structure, the product of two $G^{i8}$'s.  The flavor-$\bf 27$
mass splittings are evaluated using the relations Eq.~(\ref{t8g8}) for $T^8$
and $G^{i8}$.  For the spin-$1/2$ baryons,
\begin{eqnarray}
\P_{1 \over 2} \{ T^8, T^8 \} \P_{1 \over 2}&&=-{3 \over 2},
\nonumber\\
\P_{1 \over 2} \{T^8, \{ J^i, G^{i8}\} \P_{1 \over 2}&&=-{3 \over 2},
\end{eqnarray}
on the Gell-Mann--Okubo combination.  The product of two $G^{i8}$'s with
an intermediate spin-$1/2$ baryon
\begin{equation}
\P_{1 \over 2} G^{i8} \P_{1 \over 2} G^{i8} \P_{1 \over 2}
= -{1 \over {16}}
\end{equation}
is obtained by evaluating the operator
\begin{equation}
\P_{1 \over 2} J^i G^{i8}
\P_{1 \over 2} J^j G^{j8} \P_{1 \over 2}
={3 \over 4} \ \P_{1 \over 2} G^{i8} \P_{1 \over 2} G^{i8} \P_{1 \over 2}\ .
\end{equation}
on the Gell-Mann--Okubo combination.  The remaining operator, the product
of two $G^{i8}$'s with an intermediate spin-$3/2$ baryon is readily obtained
using the
$(0,{\bf 27})$ operator identity \cite{djmtwo}
\begin{equation}
\{ G^{i8}, G^{i8} \} = {1 \over 4} \{ T^8, T^8 \},
\end{equation}
which implies that
\begin{eqnarray}
\P_{1 \over 2} G^{i8} \P_{1 \over 2} G^{i8} \P_{1 \over 2}
+ \P_{1 \over 2} G^{i8} \P_{3 \over 2} G^{i8} \P_{1 \over 2}
&&= {1 \over 4} \P_{1 \over 2} T^8 T^8 \P_{1 \over 2}.
\end{eqnarray}
Thus,
\begin{equation}
\P_{1 \over 2} G^{i8} \P_{3 \over 2} G^{i8} \P_{1 \over 2}
= -{1 \over {8}} \ .
\end{equation}
Using these matrix elements, one obtains the nonanalytic contribution
to the Gell-Mann--Okubo mass splitting, Eq.~(\ref{gmo}).

The evaluation of the operator products for the spin-$3/2$ baryons
is similar.  For the spin-$3/2$ baryons,
\begin{eqnarray}
\P_{3 \over 2} \{ T^8, T^8 \} \P_{3 \over 2}&&=-{3},
\nonumber\\
\P_{3 \over 2} \{T^8, \{ J^i, G^{i8}\} \P_{3 \over 2}&&=-{{15} \over 2},
\end{eqnarray}
on the equal spacing rule flavor-$\bf 27$ combination.
The matrix element
\begin{equation}
\P_{3 \over 2} G^{i8} \P_{3 \over 2} G^{i8} \P_{3 \over 2} = -{5 \over 8},
\end{equation}
follows from the evaluation of
\begin{equation}
\P_{3 \over 2} J^i G^{i8}
\P_{3 \over 2} J^j G^{j8} \P_{3 \over 2}
={{15} \over 4} \P_{3 \over 2} G^{i8} \P_{3 \over 2} G^{i8} \P_{3 \over 2}\ .
\end{equation}
on the equal spacing rule mass combination.  The remaining operator is
determined using the identity
\begin{eqnarray}
\P_{3 \over 2} G^{i8} \P_{3 \over 2} G^{i8} \P_{3 \over 2}
+ \P_{3 \over 2} G^{i8} \P_{1 \over 2} G^{i8} \P_{3 \over 2}
&&= {1 \over 4} \P_{3 \over 2} T^8 T^8 \P_{3 \over 2},
\end{eqnarray}
to be
\begin{equation}
\P_{3 \over 2} G^{i8} \P_{1 \over 2} G^{i8} \P_{3 \over 2} = {1 \over 4}.
\end{equation}
Using these matrix elements, one obtains the nonanalytic contribution
to the flavor-$\bf 27$ equal spacing rule mass splitting, Eq.~(\ref{esr}).

\vfill\break\eject

\twocolumn
\narrowtext

\vfill\break\eject

\centerline{\bf Figure Captions}

\begin{figure}
\caption{$SU(2F)$ spin-flavor representation for ground-state baryons.
The Young tableau has $\N$ boxes.}
\label{fig:groundstate}
\end{figure}

\begin{figure}
\caption{$SU(F)$ flavor representations for the tower of baryon states
with $J=\frac 1 2$, $\frac 3 2$, $\ldots$, $\frac \N 2$.  Each Young
tableau has $\N$ boxes.}
\label{fig:flavorreps}
\end{figure}

\begin{figure}
\caption{Weight diagram for the $SU(3)$ flavor representation of
the spin-${1 \over 2}$ baryons. The long side of the weight diagram
contains ${1 \over 2}\left( \N + 1 \right)$ weights. The numbers
denote the multiplicity of the weights.}
\label{fig:weight1/2}
\end{figure}

\begin{figure}
\caption{Weight diagram for the $SU(3)$ flavor representation of
the spin-${3 \over 2}$ baryons. The long side of the weight diagram
contains ${1 \over 2}\left( \N - 1 \right)$ weights. The numbers
denote the multiplicity of the weights.}
\label{fig:weight3/2}
\end{figure}

\begin{figure}
\caption{Planar QCD flavor breaking at order $1/\N$ due to a
single quark loop.}
\label{fig:quarkloop}
\end{figure}

\begin{figure}
\caption{Feynman diagram responsible for
flavor-$\bf 27$
baryon mass splittings at leading order in the
flavor breaking and $1/\N$ expansions.}
\end{figure}

\vfill\break\eject

\onecolumn
\widetext

\def\sqr{\mathchoice\ssqr8{.4}\ssqr8{.4}\ssqr{5}{.3}\ssqr{4}{.3}}
\def\bsqr{\ssqr{10}{.1}}
\def\nbox{\hbox{$\bsqr\bsqr\bsqr\bsqr\raise2.7pt\hbox{$\,\cdot\cdot
\cdot\cdot\cdot\,$}\bsqr\bsqr\bsqr$}}

\centerline{$$\nbox$$}
\bigskip
\centerline{Figure 1}
\vskip1in

\def\nboxA{\vbox{\hbox{$\bsqr\bsqr\bsqr\bsqr\raise2.7pt\hbox{$\,\cdot
\cdot\cdot\cdot\cdot\,$}\bsqr\bsqr$}\nointerlineskip
\kern-.3pt\hbox{$\bsqr$}}}

\def\nboxE{\vbox{\hbox{$\bsqr\bsqr\bsqr\raise2.7pt\hbox{$\,\cdot\cdot
\cdot\cdot\cdot\,$}\bsqr\bsqr\bsqr\bsqr$}\nointerlineskip
\kern-.2pt\hbox{$\bsqr\bsqr\bsqr\raise2.7pt\hbox{$\,\cdot\cdot\cdot
\cdot\cdot\,$}\bsqr$}}}

\def\nboxF{\vbox{\hbox{$\bsqr\bsqr\bsqr\bsqr\raise2.7pt\hbox{$\,\cdot
\cdot\cdot\cdot\cdot\,$}\bsqr\bsqr$}\nointerlineskip
\kern-.2pt\hbox{$\bsqr\bsqr\bsqr\bsqr\raise2.7pt\hbox{$\,\cdot\cdot
\cdot\cdot\cdot\,$}\bsqr$}}}

\centerline{ $\nboxF$ \hskip.5in $\nboxE$ \hskip.5in
$\cdots$ \hskip.5in $\nbox$ }
\vskip.25in
\centerline{ ${J= \frac 1 2}$ \hskip1.35in
${J= \frac 3 2}$ \hskip2.35in
${J= \frac \N 2}$}
\vskip.5in
\centerline{Figure 2}
\vskip1in

\def\onedot{\makebox(0,0){$\scriptstyle 1$}}
\def\twodot{\makebox(0,0){$\scriptstyle 2$}}
\def\threedot{\makebox(0,0){$\scriptstyle 3$}}
\def\fourdot{\makebox(0,0){$\scriptstyle 4$}}

\setlength{\unitlength}{3mm}

\centerline{\hbox{
\begin{picture}(20.79,18)(-10.395,-8)
\multiput(-1.155,10)(2.31,0){2}{\onedot}
\multiput(-2.31,8)(4.62,0){2}{\onedot}
\multiput(-3.465,6)(6.93,0){2}{\onedot}
\multiput(-4.62,4)(9.24,0){2}{\onedot}
\multiput(-5.775,2)(11.55,0){2}{\onedot}
\multiput(-6.93,0)(13.86,0){2}{\onedot}
\multiput(-8.085,-2)(16.17,0){2}{\onedot}
\multiput(-9.24,-4)(18.48,0){2}{\onedot}
\multiput(-10.395,-6)(20.79,0){2}{\onedot}
\multiput(-9.24,-8)(2.31,0){9}{\onedot}
\multiput(0,8)(2.31,0){1}{\twodot}
\multiput(-1.155,6)(2.31,0){2}{\twodot}
\multiput(-2.31,4)(2.31,0){3}{\twodot}
\multiput(-3.465,2)(2.31,0){4}{\twodot}
\multiput(-4.62,0)(2.31,0){5}{\twodot}
\multiput(-5.775,-2)(2.31,0){6}{\twodot}
\multiput(-6.93,-4)(2.31,0){7}{\twodot}
\multiput(-8.085,-6)(2.31,0){8}{\twodot}
\end{picture}
}}
\bigskip
\centerline{Figure 3}
\vskip1in

\centerline{\hbox{
\begin{picture}(20.79,18)(-8.085,-8)
\multiput(-1.155,10)(2.31,0){4}{\onedot}
\multiput(-2.31,8)(9.24,0){2}{\onedot}
\multiput(-3.465,6)(11.55,0){2}{\onedot}
\multiput(-4.62,4)(13.86,0){2}{\onedot}
\multiput(-5.775,2)(16.17,0){2}{\onedot}
\multiput(-6.93,0)(18.48,0){2}{\onedot}
\multiput(-8.085,-2)(20.79,0){2}{\onedot}
\multiput(-6.93,-4)(18.48,0){2}{\onedot}
\multiput(-5.775,-6)(16.17,0){2}{\onedot}
\multiput(-4.62,-8)(2.31,0){7}{\onedot}
\multiput(0,8)(2.31,0){3}{\twodot}
\multiput(-1.155,6)(6.93,0){2}{\twodot}
\multiput(-2.31,4)(9.24,0){2}{\twodot}
\multiput(-3.465,2)(11.55,0){2}{\twodot}
\multiput(-4.62,0)(13.86,0){2}{\twodot}
\multiput(-5.775,-2)(16.17,0){2}{\twodot}
\multiput(-4.62,-4)(13.86,0){2}{\twodot}
\multiput(-3.465,-6)(2.31,0){6}{\twodot}
\multiput(1.155,6)(2.31,0){2}{\threedot}
\multiput(0,4)(4.62,0){2}{\threedot}
\multiput(-1.155,2)(6.93,0){2}{\threedot}
\multiput(-2.31,0)(9.24,0){2}{\threedot}
\multiput(-3.465,-2)(11.55,0){2}{\threedot}
\multiput(-2.31,-4)(2.31,0){5}{\threedot}
\multiput(2.31,4)(2.31,0){1}{\fourdot}
\multiput(1.155,2)(2.31,0){2}{\fourdot}
\multiput(0,0)(2.31,0){3}{\fourdot}
\multiput(-1.155,-2)(2.31,0){4}{\fourdot}
\end{picture}
}}
\bigskip
\centerline{Figure 4}
\vskip1in
\insertfig{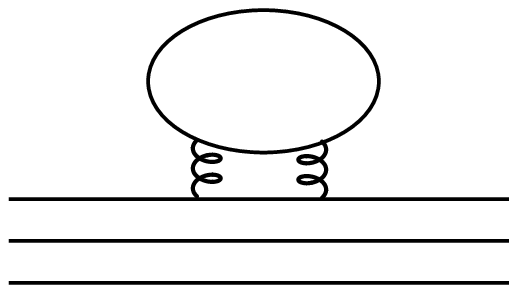}
\smallskip
\centerline{Figure 5}
\vskip1in
\insertfig{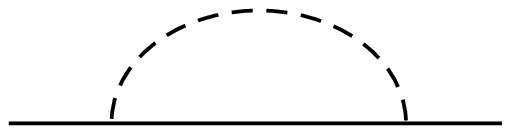}
\smallskip
\centerline{Figure 6}

\end{document}